\documentclass{jfm}

\usepackage{graphicx}
\usepackage{natbib}
\usepackage{amssymb}
\usepackage{amsmath}
\usepackage{dsfont}
\usepackage{dcolumn}

\ifCUPmtlplainloaded \else
  \checkfont{eurm10}
  \iffontfound
    \IfFileExists{upmath.sty}
      {\typeout{^^JFound AMS Euler Roman fonts on the system,
                   using the 'upmath' package.^^J}%
       \usepackage{upmath}}
      {\typeout{^^JFound AMS Euler Roman fonts on the system, but you
                   dont seem to have the}%
       \typeout{'upmath' package installed. JFM.cls can take advantage
                 of these fonts,^^Jif you use 'upmath' package.^^J}%
      }
  \else
  \fi
\fi

\ifCUPmtlplainloaded \else
  \checkfont{msam10}
  \iffontfound
    \IfFileExists{amssymb.sty}
      {\typeout{^^JFound AMS Symbol fonts on the system, using the
                'amssymb' package.^^J}%
       \usepackage{amssymb}%
         \let\leq=\leqslant
         
      }{}
  \fi
\fi

\ifCUPmtlplainloaded \else
  \IfFileExists{amsbsy.sty}
    {\typeout{^^JFound the 'amsbsy' package on the system, using it.^^J}%
     \usepackage{amsbsy}}
    {}
\fi

\newsavebox{\astrutbox}
\sbox{\astrutbox}{\rule[-5pt]{0pt}{20pt}}

\newcommand\bx{{\bf x}}                         %% Position
\newcommand\bv{{\bf V}}                         %% Velocity in Kinetic Equation  
\newcommand\bu{{\bf u}}                         %% Vector in Conservation Laws
\newcommand\bn{{\bf n}}                         %% Impact Direction Vector
\newcommand\bU{{\bf U}}                         %% Mean Velocity 
\newcommand\bF{{\bf F}}                         %% Forcing 
\newcommand\bP{{\bf P}}                         %% Pressure tensor 
\newcommand\bE{{\bf E}}                         %% E tensor
\newcommand\bq{{\bf q}}                         %% Heat flux
\newcommand\bff{{\bf f}}                        %% Vector in Conservation Laws
\newcommand\bg{{\bf g}}                         %% Vector in Conservation Laws
\newcommand\bS{{\bf S}}                         %% Vector in Conservation Laws
\newcommand\bh{{\bf h}}                         %% Local flux Vector 
\newcommand\bR{{\bf R}}                         %% Eigenvectors
\newcommand\bL{{\bf \Lambda}}                   %% Eigenvalues
\newcommand\bvv{{\bf v}}                        %% Local variables vector
\newcommand\br{{\bf r}}                         %% Position vector

\title[Pattern Formation in Vertically Oscillated Granular Disks Layers]{Granular Hydrodynamics and
  Pattern Formation in Vertically Oscillated Granular Disks Layers}

\author[J. A. Carrillo, T. P\"oschel and C. Salue\~na]%
{JOS\'E A. CARRILLO$^1$%
, THORSTEN P\"OSCHEL$^2$\break
\and CLARA SALUE\~NA$^3$}

\affiliation{$^1$ICREA (Instituci\'o Catalana de Recerca i Estudis Avan\c
cats) and Departament de Ma\-te\-m\`a\-ti\-ques, Universitat
Aut\`onoma de Barcelona, E-08193 Bellaterra, Spain\\
$^2$Charit\'e, Augustenburger Platz 1, 10439 Berlin, Germany\\
$^3$Departament de Enginyeria Mec\`anica-ETSEQ, Universitat Rovira i Virgili, E-43007 Tarragona, Spain}
\pubyear{???}
\volume{???}
\pagerange{???--???}
\date{\today\ and in revised form ??}

\begin{document}

\maketitle

\begin{abstract}
A numerical hydrodynamic model of a vibrated granular bed in 2D is
elaborated based on a highly accurate Shock Capturing scheme applied
to the compressible Navier-Stokes equations for granular flow. The
hydrodynamic simulation of granular flows is numerically difficult,
particularly in systems where dilute and dense regions occur
at the same time and interact with each other. As an example of
such a problematic system we investigate the formation of
Faraday waves in a 2d system which is exposed to vertical
vibration in the presence of gravity. The results of the CHD
agree quantitatively well with event-driven Molecular Dynamics.
\end{abstract}

\section{Introduction}

Recently, an interest has developed towards the hydrodynamic
simulation of complex granular flows, with the aim of understanding
the details of transport in the continuum limit
\citep{Goldhirsch:2003}. Being both highly nonlinear and remarkably
nonlocal, the mechanisms of granular transport in the dense limit are
still obscure. The microscopic foundation of the hydrodynamic
approaches of granular flow lies on the standard kinetic theory for
molecular gases, except for the fact that dissipation is incorporated
(via a constant or velocity dependent restitution) at the
collision. While strictly, hydrodynamic descriptions cease to be valid
near the close-packing limit, and at low restitutions, such models
have however been used successfully in situations far from their
supposed limits of validity, to describe for instance shock waves in
granular gases \citep{RBSS02,BMSS02}, clustering \citep{HM03,BSSP04}
and coexisting phases
\citep{MeersonPoeschelBromberg:2003,MeersonEtAl:2004}.  The difficulty
of a hydrodynamic description of granular materials has been addressed
in well reasoned terms in \cite{GoldhirschChaos99,LNP564}. One cannot
pretend to overcome the problems of such a description, however, where
it works exceptionally well, one should ask how, and why.

In order to compensate for the energy lost in collisions, the
typical mechanism of forcing, which is periodic vibration,
induces not only the occurrence of various interesting phenomena
but also the propagation of shock waves originating from the
moving boundary into the system. The flow developed under these
conditions is neatly supersonic and consequently, steep gradients
arise in  the hydrodynamic fields. For careful hydrodynamic simulations, it is
then desirable to use shock-capturing methods to track the sharp fronts,
 as well as high-order schemes which provide an accurate definition of the
profiles. In the literature different methods have been applied to
solve the Navier-Stokes equations for granular media, however,
most of them  lack the necessary analysis to discriminate the
effects of the implicit or artificial diffusion added to handle
supersonic flow.

In order to show the possibilities of state-of-the-art
hydrodynamic simulations (HD) applied to granular matter, in this paper
we address a paradigmatic problem of granular dynamics: the
observation of Faraday waves and the quantitative
description/comparison of the instability via both hydrodynamic
simulations based on WENO (Weighted Essentially Non-Oscillatory)
schemes and the well established method of event-driven molecular
dynamics (MD). The difficulty that a hydrodynamic code faces
when modeling such systems is twofold: the one introduced by the
instability itself, regarding the use of proper parameters and the
fine tuning of the code, and  the one mentioned above,  involving
the propagation of shock waves across the granular bed under
vibration and the discontinuities implied.

The onset of patterns in vertically oscillated granular layers has been
numerically investigated via particle and hydrodynamic simulations by
\cite{Bougieetal04}. The numerical scheme employed there to solve the
granular Navier-Stokes equations is based upon finite differences plus
artificial diffusion. In the paper a thorough analysis is done on the role of
fluctuations in the pattern inception and its dependence on the acceleration
parameter, and the results mainly refer to the mean square displacement of the
local centre of mass of the granular layer as an order parameter. Our analysis,
being also focused on the comparison between particle and hydrodynamic
simulations, is concerned with the whole set of hydrodynamic aspects related
to the instability and how they are reproduced in order to assess the validity
of granular hydrodynamic simulations in general in complex flow problems,
which is not the object of study of the mentioned paper. 

The paper is organized as follows: we start in the next section
summarizing the hydrodynamic description of rapid granular flows
based on kinetic theory. Section \ref{NS} is devoted to explain
the details of the hydrodynamic code. In Section \ref{ne} we
describe the numerical hydrodynamic experiment and the molecular
dynamics experiment; the results from both approaches are analyzed
and compared in Section \ref{Results}. In the conclusions we
summarize our findings and expose a final comparison of both
methods.

\section{\label{gkt}Granular Kinetic Theory and Hydrodynamics}

As usual in granular kinetic theory, we follow the evolution of
the number density of particles in phase space \citep{BP04}. Given
a 2D granular gas composed of small homogeneous disks of diameter
$\sigma>0$ and given $\left(\bx,\bv_1\right)$, $\left(\bx-\sigma \bn,\bv_2\right)$~the states of
two particles before the collision, where $\bn\in S^1$ is the unit
vector along the disks centers, the post-collision velocities are
found by assuming that a fraction of the normal relative velocity
is lost while its orthogonal component is unchanged. As a
consequence, the post-collision velocities are obtained as
\begin{equation}
  \begin{split}
    \bv_1^\prime &= \frac12 \left(\bv_1+\bv_2\right) + \frac{\bv^\prime}2 \\
    \bv_2^\prime &= \frac12 \left(\bv_1+\bv_2\right) - \frac{\bv^\prime}2
  \end{split}
\label{eq:Stoss}
\end{equation}
where $\bv^\prime= \bv-(1+e)\left(\bv\cdot \bn\right)\bn$, $\bv=\bv_1-\bv_2$ and $\bv^\prime =
\bv_1^\prime-\bv_2^\prime$, with $e$ being the coefficient of
(constant) restitution. Let us denote by $\bv_1^*$ and $\bv_2^*$ the
pre-collision velocities corresponding to $\bv_1$ and $\bv_2$. The
Boltzmann equation for inelastic hard disks gives the evolution of
$f\left(t,\bx,\bv_1\right)$ and it can be written as
\begin{equation}\label{boltzmann}
\frac{\partial f}{\partial t} + \left(\bv_1\cdot\nabla_\bx\right) f = \sigma^2 Q
(f,f)
\end{equation}
where the collision operator is given by
\begin{equation}
Q(f,f) = G(\nu)\! \int_{\mathds{R}^2} \int_{S^1_+}\!
\left[\left(\bv_1-\bv_2\right)\cdot \bn\right] \left[\frac1e J f\left(\bv_1^*\right)
f\left(\bv_2^*\right) - f\left(\bv_1\right) f\left(\bv_2\right)
\right]\,d\bn\,d\bv_2
\end{equation}
in which dense gas effects are taken into account through the pair
correlation function $G(\nu)$, where $\nu$ is the 2D volume
fraction, i.e., $\nu=\frac{\pi}{4}\rho\sigma^2$. The notation $S^1_+$
means that the above integral on $\bn$ is taking over those values of $\bn$
such that $(\left(\bv_1-\bv_2\right)\cdot \bn)>0$. The moments of $f$
allow us to compute the macroscopic quantities: the number
density
\begin{equation}
\rho \left(t,\bx\right) = \int_{\mathds{R}^2} f\left(t,\bx,\bv_1\right)\,d\bv_1,
\end{equation}
the velocity field
\begin{equation}
\bU (t,\bx) = \frac{1}{\rho} \int_{\mathds{R}^2} \bv_1 f\left(t,\bx,\bv_1\right)\,d\bv_1,
\end{equation}
and the granular temperature
\begin{equation}
T(t,\bx) = \frac{1}{2 \rho} \int_{\mathds{R}^2} \left|\bv_1-\bU(t,\bx)\right|^2
f\left(t,\bx,\bv_1\right)\,d\bv_1 .
\label{temperature}
\end{equation}
Different formulas are proposed for $G(\nu)$ in the literature but
we choose the accurate formulas obtained by \cite{T95}. More
precisely, we use
%% \begin{equation}
%% G(\nu)=\left\{ \begin{array}{lcr} \displaystyle \frac{1-0.436\nu}{(1-\nu)^2} &
%%     \mbox{for} & 0\leq \nu < \nu_f \\[6mm]
%% \displaystyle \frac{1-0.436\nu_f}{(1-\nu_f)^2}\;\frac{\nu_c-\nu_f}{\nu_c-\nu} &
%% \mbox{for} & 0\leq \nu_f\leq \nu_c\\
%% \end{array}\right.
%% \end{equation}
\begin{equation}
G(\nu)=
\begin{cases}
  \frac{1-0.436\nu}{(1-\nu)^2} & \mbox{for}~~~ 0\leq \nu < \nu_f \\[3mm]
  \frac{1-0.436\nu_f}{\left(1-\nu_f\right)^2}\;\frac{\nu_c-\nu_f}{\nu_c-\nu} & \mbox{for}~~~ 0\leq \nu_f\leq \nu_c\end{cases}
\end{equation}
where $\nu_f=0.69$ and $\nu_c=0.82$ are the freezing packing
fraction and the random close packing fraction respectively.

By expansion methods \citep{JR85,GS95}, the hydrodynamic equations
are obtained for the 2D granular gas,
\begin{equation}\label{ns}
\begin{split}
\frac{\partial \rho}{\partial t} + \nabla\cdot  (\rho \bU) \ &= 0 \\
\rho \left( \frac{\partial \bU}{\partial t} + (\bU\cdot \nabla)\bU \right) & = \nabla\cdot
\bP + \rho \, \bF \\
\rho \left( \frac{\partial T}{\partial t} + (\bU\cdot
\nabla)T \right)  &= -\nabla\cdot \bq + \bP:\bE - \gamma
\end{split}
\end{equation}
representing the evolution of the number density of particles
$\rho(t,\bx)$, the velocity
field $\bU(t,\bx)=\left(U_1,U_2\right)(t,\bx)$ and the granular
temperature $T(t,\bx)$. $\bP=(P_{ij})$ is the pressure tensor, given
in terms of the tensor $\bE=(E_{ij})$ by
\begin{equation}
P_{ij} = \left[ - p + \left( 2\mu_1 -  \mu_2 \right)
\sum_{i} E_{ii} \right] \delta_{ij} + 2\mu_2 E_{ij}
\end{equation}
with
\begin{equation}
E_{ij}=\frac12 \left(\frac{\partial U_i}{\partial x_j} + \frac{\partial U_j}{\partial x_i}\right)\,.
\end{equation}
The principal pressure is related to the density and the temperature through the equation of state
\begin{equation}\label{pjr}
p= \rho\epsilon \left[1+(1+e) G(\nu)\right] ,
\end{equation}
correction to the usual perfect gases law, $p=\rho\epsilon$ with
$\epsilon= T$ in 2D, proposed by \cite{JR85,GS95,BSSP04} to
incorporate dense gas effects. The vector $\bq=-\chi \nabla T$
models the heat flux. The viscosities $\mu_1$ and $\mu_2$, the
thermal conductivity $\chi$ and the cooling coefficient $\gamma$
are density and temperature dependent whose explicit expression we
take from \cite{JR85,JR852}. The bulk viscosity is given by
\begin{equation}\label{kincoef1}
\mu_1 = \frac{2}{\sqrt{\pi}} \rho \sigma T^{1/2} G(\nu)
\end{equation}
and the shear viscosity by
\begin{equation}\label{kincoef2}
\mu_2 = \frac{\sqrt{\pi}}{8} \rho \sigma T^{1/2} \left[
\frac{1}{G(\nu)} + 2 + \left(1 + \frac{8}{\pi} \right) G(\nu)
\right],
\end{equation}
the thermal conductivity is
\begin{equation}\label{kincoef3}
\chi = \frac{\sqrt{\pi}}{2} \rho \sigma T^{1/2} \left[
\frac{1}{G(\nu)} + 3 + \left(\frac94 + \frac{4}{\pi} \right)
G(\nu) \right]
\end{equation}
and the cooling coefficient is
\begin{equation}\label{kincoef4}
\gamma = \frac{4}{\sigma \sqrt{\pi}} \left(1-e^2\right) \rho T^{3/2}
G(\nu) .
\end{equation}
More involved hydrodynamic models incorporate higher order terms in the
gradients expansion \citep{Goldhirsch:2003}, and more accurate expressions for
the kinetic coefficients which include high density corrections
\citep{Garzo,Lutsko05}; in principle they can be easily implemented but it
is more illustrative to keep the approach as simple as possible. 
Finally, the vector field $\bF$ represents the external force acting
on the system, for instance gravity modulated by the piston
movement changing to the piston reference frame. The mathematical
validity of the hydrodynamic approximation in the Euler case was
discussed in \cite{BCG00}.

\section{\label{NS}Details of the hydrodynamic code}

In order to numerically simulate the granular Navier-Stokes
equations \eqref{ns}, we write them as corrections to a
compressible Euler-type system in conservation form. In fact, we
consider the equivalent system
\begin{equation}
\label{nscons}
\begin{split}
\frac{\partial \rho }{ \partial t} & + \nabla\cdot  (\rho \bU) = 0 \\
\frac{\partial (\rho \bU)}{\partial t} & + \nabla\cdot  \left[ \rho
\, (\bU \otimes \bU)\right] + \nabla p = \nabla\cdot \bP + \rho \, \bF \\
\frac {\partial W }{\partial t} & + \nabla\cdot \left[ \bU W \right]
= -\nabla\cdot \bq + \bP:\bE + \bU \cdot (\nabla\cdot \bP)- \gamma + \rho (\bU
\cdot \bF),
\end{split}
\end{equation}
where the total energy $W$ is given by 
\begin{equation}
W=\rho T + \frac12
\rho |\bU|^2=\rho \epsilon + \frac12 \rho |\bU|^2\,.
\end{equation} 
The system
\eqref{nscons} can be rewritten as
\begin{equation}
  \label{genconslaw}
  \frac{\partial \bu}{\partial t} + \frac{\partial}{\partial x_1}
\bff\left(\vec{u}\right) + \frac{\partial}{\partial x_2}  \bg\left(\bu\right) = \bS\left(\bu\right)
\end{equation}
with $\bx=(x_1,x_2)$ and  obvious definitions for $\bff\left(\bu\right)$,
$\bg\left(\bu\right)$ and $\bS\left(\bu\right)$ and with the
vector $\bu = \left(u_1,u_2,u_3,u_4\right)$ whose components
are given by
\begin{equation}
  \begin{split}
    u_1 &=\rho\\ %\qquad , \qquad
    u_2 &=\rho U_1\\% \qquad , \qquad
    u_3 &=\rho U_2\\
    u_4 &=\rho \epsilon + \frac12 \rho \left|\bU\right|^2=\rho \epsilon + \frac12 \rho \left(U_1^2+U_2^2\right)\,.
  \end{split}
\end{equation}
As a consequence, equations \eqref{nscons} have the structure of a system of
nonlinear conservation laws with sources. Local eigenvalues and
both local left- and right-eigenvectors of the Jacobian matrices
$\bff^\prime\left(\bu\right)$ and $\bg^\prime\left(\bu\right)$
are explicitly computable and included in the appendix for the
sake of the reader. Since local eigenvalues and eigenvectors need
the derivative of the function $G(\nu)$ appearing in the equation
of state \eqref{pjr}, then $G(\nu)$ is smoothed out around the
freezing point volume fraction $\nu_f$.

The numerical method consists in applying a high-order
shock-capturing scheme for the Euler part of this system, i.e.,
the left-side of \eqref{genconslaw}, while the terms in the
right-hand side $\bS\left(\bu\right)$ are approximated by second
order central finite differences.

The convective terms $\frac{\partial}{\partial x_1}
\bff\left(\bu\right)$ and $\frac{\partial}{\partial x_2}
\bg\left(\bu\right)$ are approximated by a fifth-order finite
difference characteristic-wise WENO method in a uniform grid
following \cite{JS96}. This scheme is a well-known shock-capturing
and high-order method for nonlinear conservation laws. Extensive
benchmarks have shown this method to be particularly well adapted
to control oscillations in shock regimes in classical Euler
equations for gases, see the survey by \cite{Shu98}.

We first restrict ourselves to the one dimensional procedure,
working in a dimension-by-dimension fashion for reconstructing
each of the fluxes in Eq. \eqref{genconslaw}. This means that when
approximating $\frac{\partial }{\partial x_1}\left(
\bff\left(\bu\right)\right)$, for example, the other 
variable $x_2$ is fixed and the approximation is
performed along the $x_1$ line. Given a uniform grid $r_i$ for any
of the spatial variables denoted by $r$, let us call
$\bu_i(t)$ the numerical approximation to the point value
$\bu\left(r_i,t\right)$. The corresponding convective term is
approximated as
\begin{equation}
\frac{\partial}{\partial r} \bff\left(\bu\right) \ \simeq \
\frac{\hat{\bff}_{i+\frac12} - \hat{\bff}_{i-\frac12}}{\Delta r}
\end{equation}
where $\hat{\bff}_{i+\frac12}$ is the numerical flux.

Let us briefly describe the general characteristic-wise finite
difference procedure with flux splitting and flux reconstruction.
At each fixed time $t$, we find the numerical fluxes
$\hat{\bff}_{i+\frac12}$ in the following way:
\begin{itemize}
\item[\bf{Step 1.}] We first compute an average or intermediate
state $\bu_{i+\frac12}$ using Roe's formula as in
\cite{Shu98,KGSD00}.

\item[\bf{Step 2.}] We compute the matrices of right $\bR$ and left
  $\bR^{-1}$ eigenvectors and eigenvalues $\Lambda_l\left(\bu_{i+\frac12}\right)$,
  $l=1,\dots 4$, of the Jacobian
  matrix $\bff^\prime\left(\bu_{i+\frac12}\right)$ from the formulas postponed to the appendix to obtain
\begin{equation}
  \begin{split}
    \bR&=\bR\left(\bu_{i+\frac12}\right)\\% \quad , \quad
    \bR^{-1}&=\bR^{-1}\left(\bu_{i+\frac12}\right)\\% \quad \mbox{and} \quad 
    \bL&=\bL\left(\bu_{i+\frac12}\right)=\mbox{diag}\left[\Lambda_l\left(\bu_{i+\frac12}\right)\right]\,.
  \end{split}
\end{equation}

\item[\bf{Step 3.}] We transform the values $\bu_i$ and
$\bff\left(\bu_i\right)$ in the potential stencil of the approximation of
the flux to the local characteristic coordinates by using the left
eigenvectors:
\begin{equation}
\bvv_i = \bR^{-1} \bu_i \quad \mbox{and} \quad \bh_i = \bR^{-1} \bff\left(\bu_i\right).
\end{equation}

\item[\bf{Step 4.}] We use global Lax-Friedrichs flux splitting
with the transport coefficient computed from the maximum, in
absolute value, among the above computed eigenvalues to get
$\bh^\pm_i$ from
\begin{equation}
\bh^\pm_i = \frac12 \left(\bh_i \pm \alpha \bvv_i\right),
\end{equation}
where the dissipation parameter $\alpha$ is given by
\begin{equation}
\alpha = \max_{i,l} \left|\Lambda_l\left(\bu_{i+\frac12}\right)\right|.
\end{equation}

\item[\bf{Step 5.}] We compute the decomposed fluxes
$\hat{\bh}_{i+\frac12}^+$ and $\hat{\bh}_{i+\frac12}^-$ at the middle
points using a high-order reconstruction explained below.

\item[\bf{Step 6.}] We transform back the computed fluxes to the
physical space by using the right eigenvectors 
\begin{equation}
\hat{\bff}_{i+\frac12}^\pm=\bR \hat{\bh}_{i+\frac12}^\pm
\end{equation}
and finally, we add them to obtain the final numerical flux
\begin{equation}
\hat{\bff}_{i+\frac12}=\hat{\bff}_{i+\frac12}^+ +
\hat{\bff}_{i+\frac12}^-\,.
\end{equation}
\end{itemize}
Instead of Roe mean and Lax-Friedrichs flux splitting formulas, we
can use more sophisticated and improved methods avoiding numerical
diffusion but in this case we do not need them since Navier-Stokes
terms will have a regularizing effect anyhow on shock waves in the
flow.

In the WENO5 method, reconstructions of the fluxes in Step 5 are
done using a nonlinear nonlocal convex combination of three
different approximations of the flux by three different local
stencils using the upwinding of the flux. The contribution of each
approximation for the computation of the final value of the flux
depend on the local smoothness of the function measured by certain
smoothness indicators based on the local divided differences. If
the function is smooth the resulting approximation is fifth order.
The idea behind this weighted approximation is that those stencils
including a discontinuity or high gradient will receive almost
zero weight in the approximation and thus, oscillations will be
avoided while keeping high-order approximation in smooth parts of
the flow.

More precisely, let us denote by $\hat{h}_{i+1/2,l}^+$ the $l$-th
component of the local numerical flux 
$\hat{\bh}^+_{i+1/2}$ and by  $h_{i,l}^+$ the $l$-th component of the
local flux $\bh^+_i$. Since the computations below are analogous component
by component and fixed for the chosen upwinding, we will avoid the sub-
and superindex $_{,l}^+$ for notational simplicity. Then,
we obtain the numerical flux $\hat{h}_{i+1/2}$ by
\begin{equation}
\hat{h}_{i+1/2} = \omega_1 \hat{h}_{i+1/2}^{(1)} + \omega_2
\hat{h}_{i+1/2}^{(2)} + \omega_3 \hat{h}_{i+1/2}^{(3)}
\end{equation}
where $\hat{h}_{i+1/2}^{(m)}$ are the three third order fluxes on
three different stencils given by
\begin{equation}
\begin{split}
\hat{h}_{i+1/2}^{(1)} & = \frac{1}{3} h_{i-2} - \frac{7}{6} h_{i-1}
               + \frac{11}{6} h_{i}, \\
\hat{h}_{i+1/2}^{(2)} & = -\frac{1}{6} h_{i-1} + \frac{5}{6} h_{i}
               + \frac{1}{3} h_{i+1}, \\
\hat{h}_{i+1/2}^{(3)} & = \frac{1}{3} h_{i} + \frac{5}{6} h_{i+1}
               - \frac{1}{6} h_{i+2},
\end{split}
\end{equation}
and the nonlinear weights $\omega_m$ are given by
\begin{equation}
\begin{split}
  \omega_m & = \frac {\tilde{\omega}_m}{\sum_{l=1}^3 \tilde{\omega}_l} \\ %,\qquad
  \tilde{\omega}_l & = \frac {\gamma_l}{\left(\varepsilon + \beta_l\right)^2} % ,
\end{split}
\end{equation}
with the linear weights $\gamma_l$ given by
\begin{equation}
\gamma_1=\frac{1}{10}, \qquad \gamma_2=\frac{3}{5}, \qquad
\gamma_3=\frac{3}{10},
\end{equation}
and the smoothness indicators $\beta_l$ given by
\begin{equation}
\begin{split}
\beta_1 & = \frac{13}{12} \left( h_{i-2} - 2 h_{i-1}
                             + h_{i} \right)^2 +
         \frac{1}{4} \left( h_{i-2} - 4 h_{i-1}
                             + 3 h_{i} \right)^2  \\
\beta_2 & = \frac{13}{12} \left( h_{i-1} - 2 h_{i}
                             + h_{i+1} \right)^2 +
         \frac{1}{4} \left( h_{i-1}
                             -  h_{i+1} \right)^2  \\
\beta_3 & = \frac{13}{12} \left( h_{i} - 2 h_{i+1}
                             + h_{i+2} \right)^2 +
         \frac{1}{4} \left( 3 h_{i} - 4 h_{i+1}
                             + h_{i+2} \right)^2 .
\end{split}
\end{equation}
Finally, $\varepsilon$ is a parameter to avoid the denominator to
become 0 and is taken as $\varepsilon = 10^{-6}$ in the
computation of this paper. The computation for approximating
$\hat{\bh}^-_{i+1/2}$ is obtained with the alternate upwinding by mirror
symmetry with respect to the index $i+\frac12$. We refer to the
survey by \cite{Shu98} for further details about WENO
reconstruction procedures, smoothness indicators, benchmarks and
references for different applications.

Finally, the resulting ODE system is solved in time by a 3rd-order
TVD Runge-Kutta scheme introduced by \cite{SO88} imposing a local
CFL condition at every time step, ensuring that the numerical
speed is not less than the the largest eigenvalue of the local
Jacobian matrix. This CFL condition becomes more restrictive as we
approach the close-packing volume fraction. Although viscosity and
thermal conductivity terms are treated as sources, they are taken
into account in the CFL condition since diffusion effectively
paces the time stepping of the code. This fact is due to the
dependence of the viscosity and thermal conductivity on the volume
fraction, resulting in diverging coefficients for both
close-packing and vacuum limits. Both limits are avoided in the
actual computation shown in this paper by establishing a limit
from above and below of the volume fraction. These limits are
10$^{-4}$ and 99.99\% of the maximal volume fraction.

\section{\label{ne}Numerical experiments}

\subsection{Structures in vertically oscillated granular systems}

When granular matter is exposed to vertical oscillations in the
presence of gravity, one observes spatio-temporal structures at
the free surface. Although this observation was reported as early
as 1831 by \citeauthor{Faraday:1831} the effect was studied only
in recent years systematically and a variety of patterns was
observed, e.g. by \cite{PakBehringer:1993,VanDoornBehringer:1997a,MeloUmbanhowarSwinney:1994,MeloUmbanhowarSwinney:1995,MeloUmbanhowarSwinney:1996,UmbanhowarMeloSwinney:1998,MetcalfKnightJaeger:1997}
and theoretically explained, e.g., by \cite{Rothman:1998,TsimringAranson:1997}.

Here we focus on subharmonic instabilities in a vertically
vibrated layer of granular matter, first reported by
\cite{DouadyFauveLaroche:1989} and further characterized in
dependence on the parameters of the vibration by
\cite{ClementVanelRajchenbachDuran:1996,WassgrenBrennenHunt:1996}.
Above a certain amplitude of acceleration, $\Gamma=A\omega^2/g$
($A$ is the amplitude, $\omega$ the frequency of the sinusoidal
excitation, $g$ is gravity), a periodic wave-like pattern appears
with frequency $\omega/2$. Although the condition $\Gamma>1$ is
not a necessary condition for pattern formation
\citep{PoeschelSchwagerSaluena:2000,RenardEtAl:2001}, in all
experimental observations the onset of the pattern formation was
observed for $\Gamma\gtrsim 4$.

\begin{figure}
%  \newcounter{fnum}
%  \setcounter{fnum}{6}
%  \newcounter{ffnum}
%  \setcounter{ffnum}{61}
%  \newcounter{flabel}
%  \setcounter{flabel}{1}
%  \newcounter{flabelr}
%  \setcounter{flabelr}{6}
%  \def\dfnum{11}
%  \def\pif{
%    \centerline{
%      \begin{tabular}[t]{llll}
%    \theflabel & \includegraphics[width=0.45\textwidth,bb=775 1 1594 188,clip]{FIGS/snaps/\thefnum.eps} &
%    \includegraphics[width=0.45\textwidth,bb=775 1 1594 188,clip]{FIGS/snaps/\theffnum.eps} & \theflabelr
%      \end{tabular}
%    }
%    \addtocounter{fnum}{\dfnum}
%    \addtocounter{ffnum}{\dfnum}
%    \stepcounter{flabel}
%    \stepcounter{flabelr}
%  }
%  \pif \pif \pif \pif \pif
\includegraphics[width=0.99\textwidth,clip]{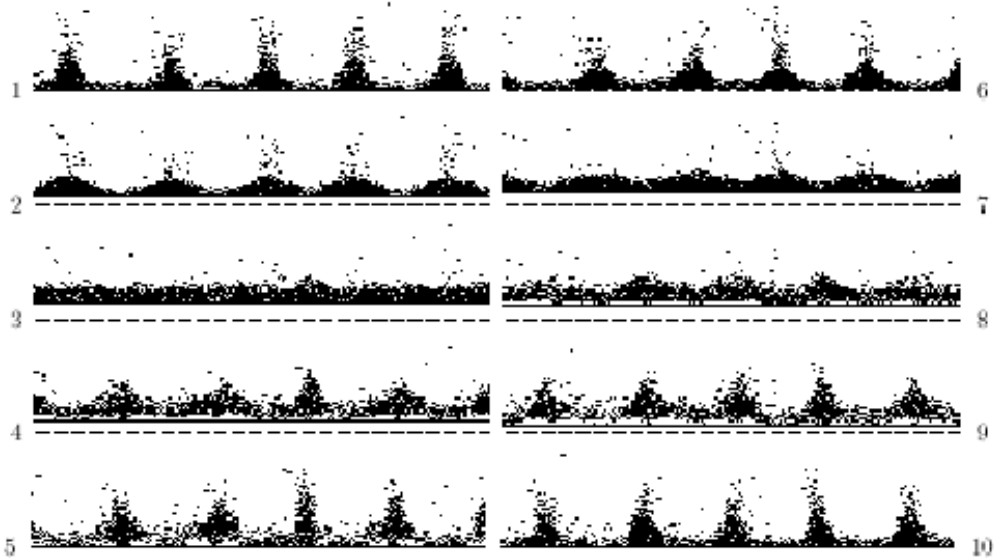}
\caption{Periodic wave-like pattern in a vertically vibrated
granular layer. The period of the pattern is twice the period of
the excitation. The figure shows a two-dimensional MD simulation
of 3000 particles at $A=4.02$, $f=3.5$ (details see below). Only a
part of the system is shown. The dashed line indicates the lower
amplitude of the oscillation. An animated sequence is provided as supplementary material.}
  \label{fig:snaps}
\end{figure}

These waves could be reproduced in quantitative agreement with
experiments by numerical MD simulations by
\cite{LudingClementRajchenbachDuran:1996,Luding:1997} and by
\cite{AokiAkiyama:1996} \citep[see
also][]{BizonShattuckNewmanUmbanhowarSwiftMcCormickSwinney:1997,AokiAkiyama:1997}
and were also found from a linear stability analysis of an
oscillating granular layer, modeled as an isothermal
incompressible fluid with vanishing surface tension
\citep{BizonShattuckSwift:1999}. Figure \ref{fig:snaps} shows
snapshots of a Molecular Dynamics simulation.

%While MD simulations could reproduce the pattern with very good
%accuracy, so far this system was not studied by Computational Fluid
%Mechanics. 
A computational fluid mechanics simulation of such a system represents a hard
problem since the granular material occurs in dense state (e.g. snapshots 
2 and 3 in Figure \ref{fig:snaps}) and in dilute state (e.g.
snapshots 4 and 9 in Figure \ref{fig:snaps}). A universal CFD
computation should be capable to describe the system in both
states dense and dilute.

The described effect and its dependence on the system parameters
was well studied in the literature; in the present paper it serves
as an example showing that our CFD simulation is in quantitative
agreement with MD simulations and, thus, capable to describe
time-dependent behavior of granular systems in dense and dilute
state. For simplicity and better visual representation, here we
restrict ourselves to a two-dimensional system, the generalization
to 3D is conceptually simple although computationally expensive.

\subsection{Setup of the Molecular Dynamics system}

Our reference system is an event-driven Molecular Dynamics
simulation as shown in Figure \ref{fig:snaps}. In event-driven MD,
each of the $N$ particles of the system moves along a parabolic
trajectory due to gravity $g$. This motion is interrupted by
binary collisions, where each of the collision partners changes
its velocity according to Eq. \eqref{eq:Stoss}. Given particle
positions and velocity at time $t$, the time of the next collision
in the entire system can be computed analytically, therefore, in
event-driven simulations, the time progresses in irregular steps,
from event to event, i.e., from collision to the next collision.
Unlike force-based MD, where Newton's equation of motion is solved
numerically by evaluating the interaction forces, event-driven MD
does not solve Newton's equation explicitly. Instead, Newton's
equation is implied in the coefficient of restitution which
relates the relative particle velocity after a collision to the
velocity before the collision, Eq. \eqref{eq:Stoss}, that is, the
coefficient of restitution itself which, in general, is a function
of material parameters and impact velocity, is a result of solving
Newton's equation
\citep{BrilliantovEtAl:1996,SchwagerPoeschel:1998,RamirezEtAl:1999}.
In the present paper, we assume that the coefficient of
restitution is a material constant and adopts the value
$e=0.75$.

\begin{table}
\caption{Performed MD simulations. $N$ -- number of particles,
$f=2\pi\omega$ -- frequency of the oscillating bottom wall (in Hz), $A$ --
amplitude of the oscillation, $L$ -- system width, $H$ -- number
of particle layers. The observed patterns are characterized by
$N_p$ -- number of peaks, $\lambda$ -- wave length, $d\lambda$ -- standard
deviation of the measured wavelength over 50 cycles.
All lengths are given in units of the particle diameter.}
  \vspace*{0.2cm}
  \newcolumntype{d}[1]{D{.}{.}{#1}}
  \begin{center}
    \begin{tabular}{d{2}@{~~}|@{~~}d{0}@{~~}|@{~~}d{1}@{~~}|@{~~}d{2}@{~~}|d{3}|d{4}|d{0}|d{0}|d{1}|d{3}|d{2}}
      \Gamma & N & f & A & L & H & N_p & \lambda & d\lambda  \\
      \hline
      %g       N        f        A        L      H     np      lambda  ld
      2.00  &  1879  &  5.5  &   1.64  &  300  &  6  &  15  &  18  &  21 \\
      2.00  &  2506  &  5.0  &   1.99  &  400  &  6  &  16  &  25  &  18 \\
      2.00  &  2506  &  4.5  &   2.45  &  400  &  6  &  13  &  31  &  18 \\
      2.00  &  2506  &  4.0  &   3.12  &  400  &  6  &  11  &  37  &  15 \\
      2.00  &  3132  &  3.5  &   4.02  &  500  &  6  &  12  &  42  &  5 \\
      2.00  &  3759  &  3.0  &   5.50  &  600  &  6  &  12  &  50  &  10 \\
      \hline
      %g       N        f        A        L      H     np      lambda  ld
      2.25  &  1566 &   6.0  &   1.55  &  250  &  6  &  12  &  21  &  9  \\
      2.25  &  1879 &   5.5  &   1.85  &  300  &  6  &  13  &  23  &  7  \\
      2.25  &  2506 &   5.0  &   2.24  &  400  &  6  &  15  &  27  &  4.5  \\
      2.25  &  2506 &   4.5  &   2.76  &  400  &  6  &  14  &  29  &  4.5 \\
      2.25  &  3759 &   4.0  &   3.51  &  600  &  6  &  19  &  31  &  2 \\
      2.25  &  3759 &   3.5  &   4.52  &  600  &  6  &  14  &  43  &  7  \\
      2.25  &  3759 &   3.0  &   6.19  &  600  &  6  &  12  &  50  &  13 \\
      \hline
      %g       N        f        A        L      H     np      lambda  ld
%      2.52  &  1880  &  6.0  &   1.74  &  300  &  6  &  ?   &  20  &   12 \\
      2.52  &  1880  &  5.5  &   2.07  &  300  &  6  &  15  &  20  &   9 \\
      2.52  &  1880  &  5.0  &   2.50  &  300  &  6  &  12  &  25  &   7 \\
      2.52  &  1880  &  4.5  &   3.09  &  300  &  6  &  11  &  27  &   6 \\
      2.52  &  1880  &  4.0  &   3.90  &  300  &  6  &   9  &  33  &   6 \\
      2.52  &  3132  &  3.5  &   5.07  &  500  &  6  &  12  &  42  &   7 \\
      2.52  &  3132  &  3.0  &   6.92  &  500  &  6  &  10  &  50  &   16 \\
      \hline
      %g       N        f        A        L      H     np      lambda  ld
      2.75  &  3132  &  3.5  &   5.57  &  500  &  6  &  13  &  39  &   8 \\
      2.75  &  3132  &  3.0  &   7.59  &  500  &  6  &   9  &  56  &   13\\
      2.75  &  3759  &  2.5  &   10.9  &  600  &  6  &   9  &  67  &   15

    \end{tabular}
  \end{center}
  \label{tab:parameters}
\end{table}

The main precondition for applying event-driven MD is the
assumption of binary collisions which implies hard particles.
Although conceptually simple, an efficient implementation of
event-driven MD algorithms is by far more complex than force-based
algorithms \citep[see, e.g.,][]{PoeschelSchwager:2005}. To compare
particle dynamics with hydrodynamics, it is necessary to employ
event-driven MD since both MD and HD simulations are based on the same
equations \eqref{eq:Stoss} using both the concept of binary
collisions and the coefficient of restitution. In general, there
is no simple correspondence between force-based MD and
hydrodynamics.

The disadvantage of event-driven MD is its restriction to binary
and thus, instantaneous collisions. While for dilute granular gases
this assumption is well justified, for systems with gravity, for
mathematical reasons we cannot exclude tree-particle interaction,
also called {\em inelastic collapse}
\citep{McNamaraYoung:1993PRE}. There are numerical tricks to avoid
or retard the inelastic collapse in MD simulations
\citep[e.g.,][]{LudingNamara:1998}, provided density is small
enough, however, for systems with gravity and reflecting
oscillating bottom wall, at low frequency event-driven simulations
become problematic: Consider a ball falling vertically from a
certain height. At each collision with the wall its energy is
reduced by the factor $e^2$. Then, in {\em finite} time
the ball comes to rest, that is, it stays in permanent contact
with the wall which violates the condition of instantaneous
contacts. The same argument applies for slowly oscillating walls,
therefore, our simulations fail for low oscillation frequency.

To compare event-driven simulations with CFD results, we
perform simulations of an assembly of $N$ particles of diameter 
$\sigma=10$\,mm, colliding due to Eq. \eqref{eq:Stoss} with
coefficient of restitution $e=0.75$. The width of the
system is $L$. After the transient time when the pattern is
time-invariant (except for its period), we recorded the horizontal
positions of all particles for 50 periods. The number of
peaks $N_p$ of this pattern was then determined by visually
inspecting the histogram of this data. In addition, we determined
the wave lengths $\lambda$ of the pattern by Fourier-analysis,
which agrees well with the values for $N_p$. 

Table \ref{tab:parameters} shows the parameters of the performed
event-driven MD simulations and the results for $N_p$ and
$\lambda$, extracted from Fourier analysis. The standard deviations $d\lambda$
account for the spread of the wavelength, which at very low amplitudes /
accelerations is specially high. The range of amplitudes $A$ and accelerations
$\Gamma$ investigated is devised to capture the development of the
instability; the width of the system $L$ is chosen to ensure that a sufficient
number of wavelengths fits the simulation window -- and thus the errors in
the determination of $\lambda$ are diminished. These results will be discussed
below in Sec. \ref{Results} in  comparison with the results of the
computational hydrodynamics.  

Hydrodynamic fields such as density, velocity and temperature were
 generated from  50 cycles of the MD simulation at amplitude $A$ = 5.6$\,\sigma$
 ($\sigma$: particle diameter) and frequency $f$ = 3.5 Hz, with the aim  to
 compare  them  with the respective fields from the HD simulation. For this
 purpose we  used a $150\,\sigma\times50\,\sigma$ grid. The positions and
 velocities of all particles were recorded for 50 periods at a rate of 50
 fixed equidistant phases per period. In order to generate hydrodynamic
 fields, phase averaging was performed  over particle positions and
 velocities  with a 
 spatial coarsegraining function \citep{GoldhirschChaos99}  $\phi(\br) \propto
 \theta\left(\sigma/2-\left|\br-\br_i(t)\right|\right)$  where  $\theta$ is the Heaviside
 step  function and $\br_i(t)$ the instantaneous location of the center of
 a  particle. The cell size for averaging fits about $3\times1$
 particles, which leaves us with a typically mesoscopic procedure. 
  
\subsection{Setup for the hydrodynamic system}

We solve the full set of equations in 2D \eqref{ns} supplied with
the kinetic coefficients given by
Eqs. \eqref{kincoef1}-\eqref{kincoef4} and the equation of state
\eqref{pjr} on a $150\times(50\mbox{ up to }100)$  rectangular grid. The initial state
is prepared by leaving the material settle down under the action
of gravity to form a deposited layer of 6 grains. The depth of the
granular bed is imposed by assuming a close random packing density
in the deposited layer. As in MD simulations, the diameter of the grains is
$\sigma=10$ mm and their restitution  $e=0.75$.

The domain in which the hydrodynamic equations are solved varies
in the range of 300 to 600$\sigma$ along the $x_1$ direction, and
from 30 to 100$\sigma$ along the vertical or $x_2$ direction for the
range of amplitudes $A$ and frequencies $f$ sampled in the
simulations (1.5 to 15$\sigma$ in amplitude, 2 to 6 Hz in
frequency). This way we make sure that the scale  of the pattern
is correctly modeled and a sufficient number of pattern
wavelengths is captured. The choice of the vibration parameters is
adjusted  to cover the bifurcation region, that is, the region
where the granular layer destabilizes from the homogeneous (along
the $x_1$ direction) solution. In general, we use exactly the same values for
the parameters as in the MD simulations. 

We use periodic boundary conditions along the $x_1$ direction,
while at the top and the bottom we set reflecting boundaries. The top wall is
set at a distance where reflections are of minor importance (the packing 
fraction at those heights is always of the order of 10$^{-4}$). The
hydrodynamic fields (density, velocity components, temperature)
are stored for subsequent inspection without any additional
treatment. Since the homogeneous solution tends to be very stable
in the absence of random sources, we numerically perturb the layer
density with a collection of modes (the first 20 with largest
wavelengths), with random phases and amplitudes. The latter  are
selected  with a maximum in the order of $10^{-3}$ times the value
of the local packing fraction of the deposited material. This
small perturbation is sufficient to quickly observe the growth of
the mode selected by the system which gives rise to the
development of the Faraday waves \citep[see][]{Movie}.

\section{\label{Results}Results}

\begin{figure}
\centerline{\includegraphics[angle=0,width=0.48\textwidth,clip]{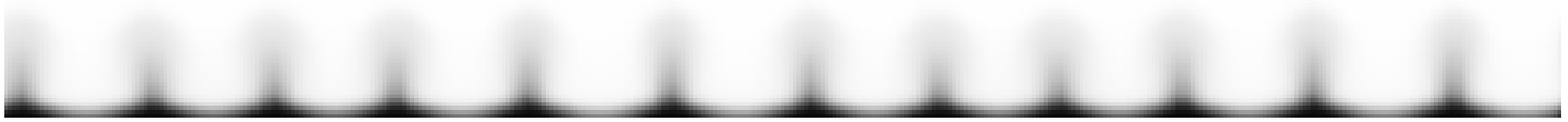}\hfill\includegraphics[angle=0,width=0.48\textwidth,bb=15 14 446 60,clip]{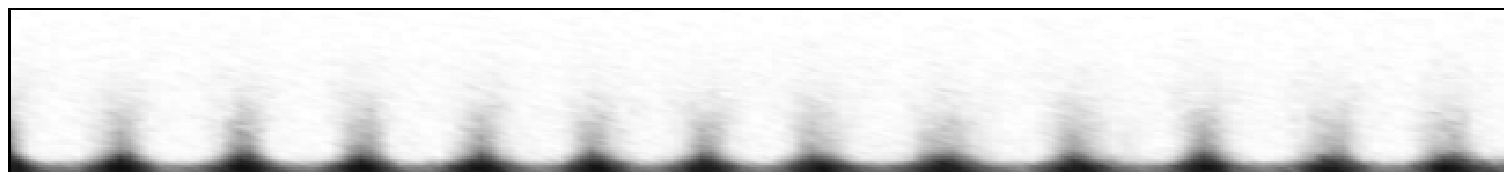}}
\centerline{\includegraphics[angle=0,width=0.48\textwidth,clip]{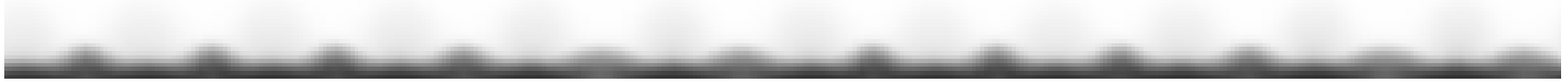}\hfill\includegraphics[angle=0,width=0.48\textwidth,bb=16 14 426 60,clip]{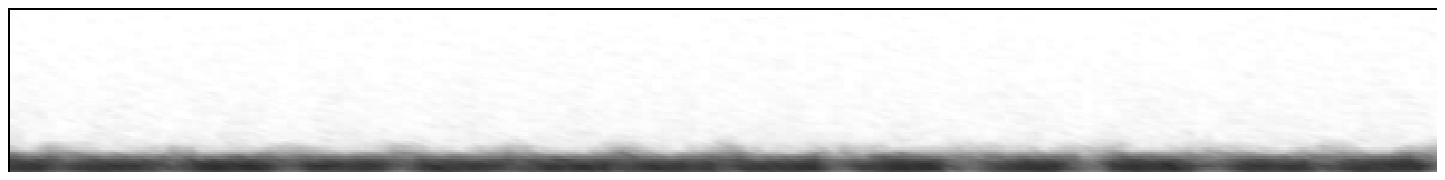}}
\centerline{\includegraphics[angle=0,width=0.48\textwidth,clip]{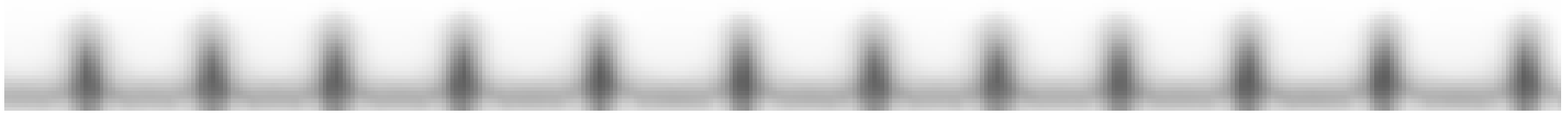}\hfill\includegraphics[angle=0,width=0.48\textwidth,bb=16 14 426 60,clip]{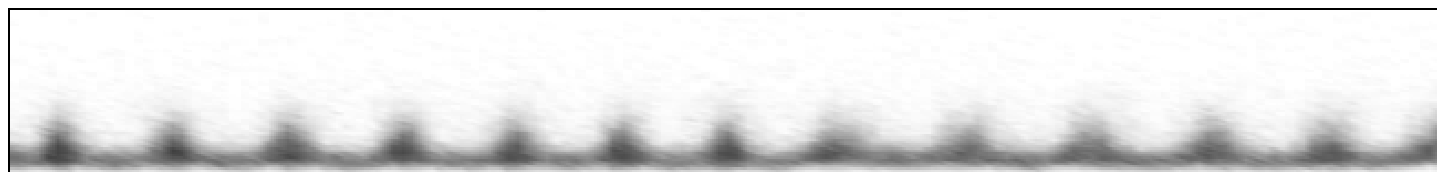}}
\centerline{\includegraphics[angle=0,width=0.48\textwidth,clip]{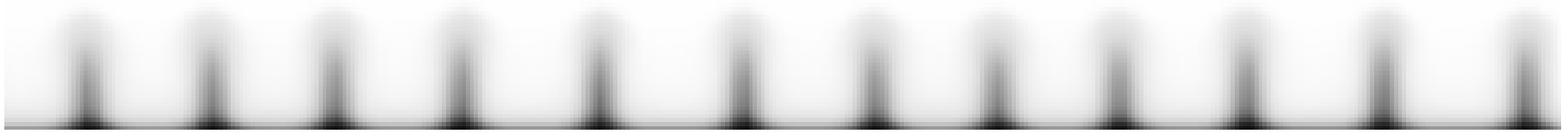}\hfill\includegraphics[angle=0,width=0.48\textwidth,bb=16 14 426 60,clip]{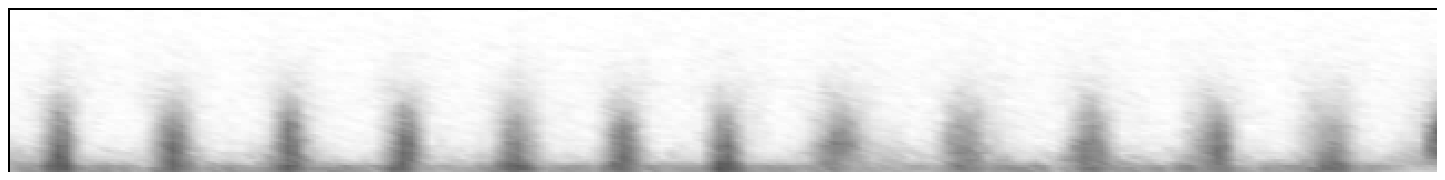}}
\centerline{\includegraphics[angle=0,width=0.48\textwidth,clip]{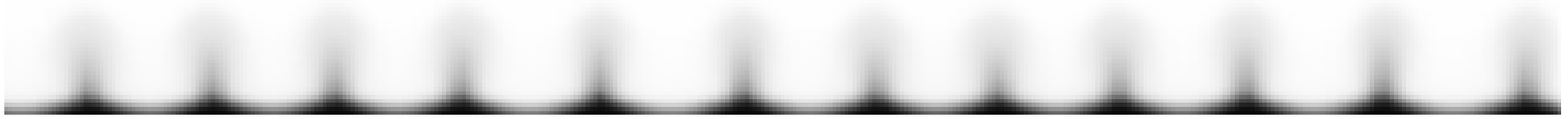}\hfill\includegraphics[angle=0,width=0.48\textwidth,bb=16 14 426 60,clip]{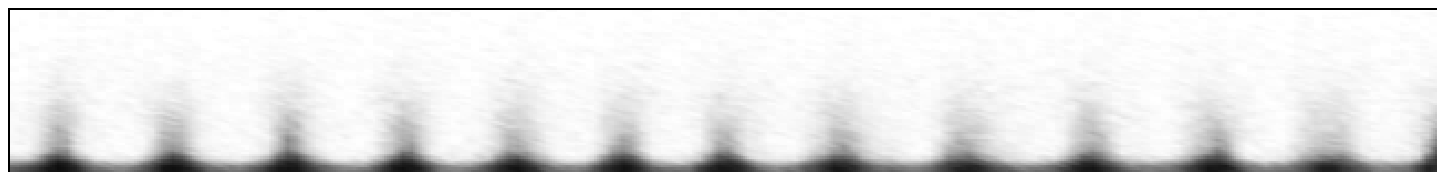}}
\centerline{\includegraphics[angle=0,width=0.48\textwidth,clip]{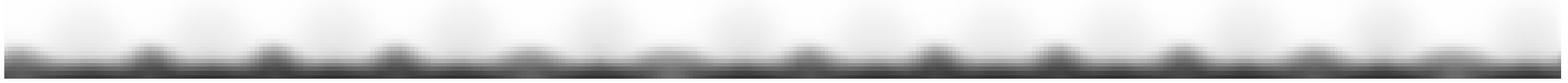}\hfill\includegraphics[angle=0,width=0.48\textwidth,bb=16 14 426 60,clip]{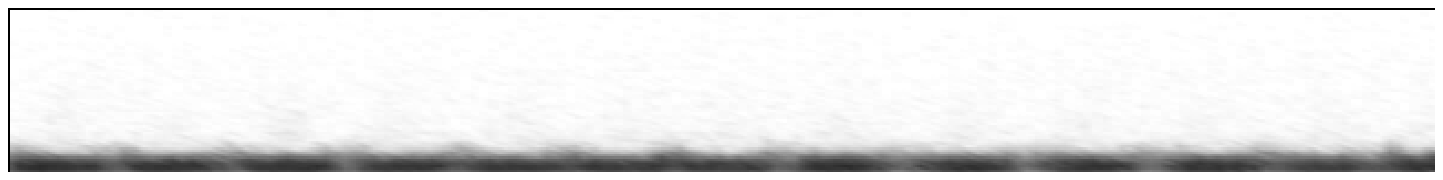}}
\centerline{\includegraphics[angle=0,width=0.48\textwidth,clip]{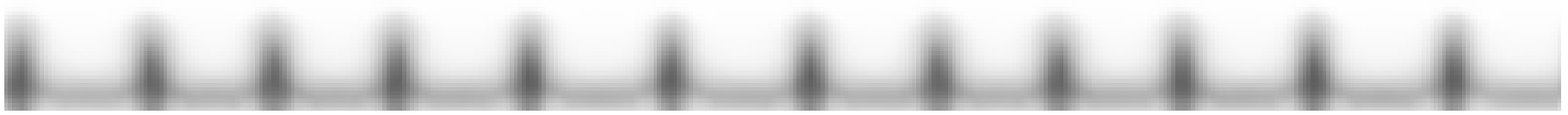}\hfill\includegraphics[angle=0,width=0.48\textwidth,bb=16 14 426 60,clip]{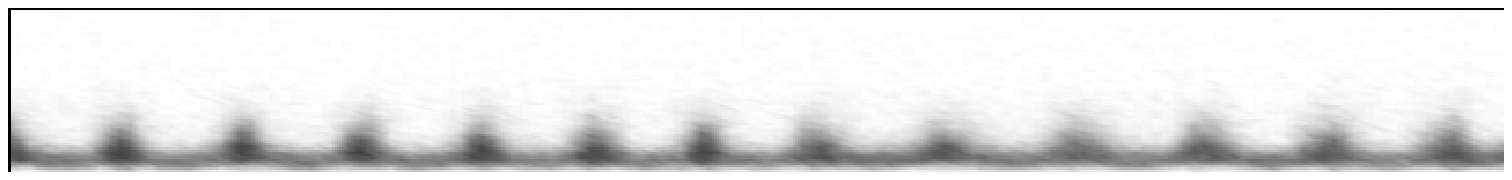}}
\centerline{\includegraphics[angle=0,width=0.48\textwidth,clip]{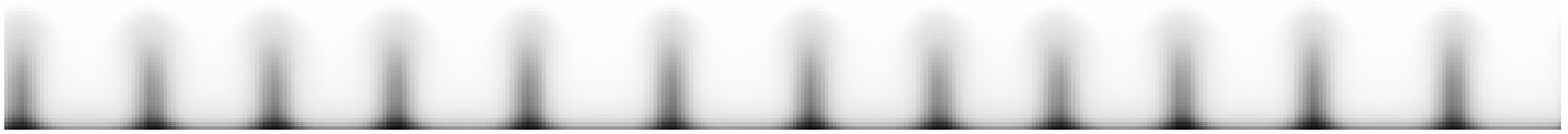}\hfill\includegraphics[angle=0,width=0.48\textwidth,bb=16 14 426 60,clip]{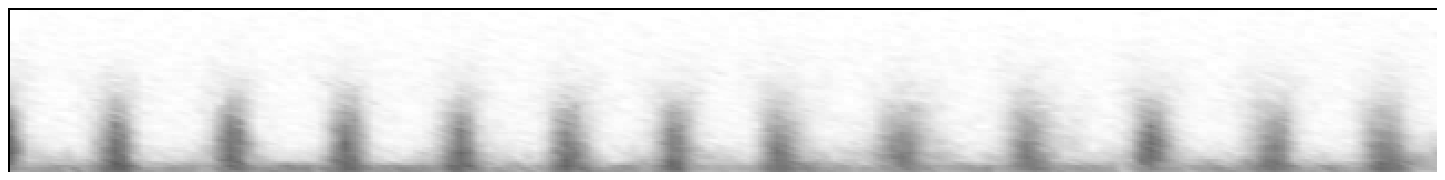}}
\caption{Eight pictures of the density as obtained from the
numerical solution of the hydrodynamic equations (HD, left) and the
corresponding ones from the molecular dynamics simulation (MD, right).
The times for the HD snapshots are: 0, $0.25T$, $0.5T$, $0.75T$, $T$, $1.25T$,
$1.5T$, $1.75T$, with $T$ being the period of the oscillation.  The MD
snapshots were taken at times: 0, $0.25T$, $0.5T$, $0.86T$, $T$, $1.25T$,
$1.5T$, $1.86T$. The initial phase is arbitrary, but  common for both
sets. For most of the period, the dynamics observed in HD and MD 
simulations is completely equivalent; the difference in the fourth and eighth
phases is due to different landing times of the granular bed (see
text). The MD figures are obtained after averaging particle positions 
over 50 cycles, while the density field on the left is the raw
output. The frequency of vibration is $f=3.5$ Hz and the amplitude
$A=5.6$ diameters. Complete sequences are available as supplementary material.} 
\label{snaps}
\end{figure}

Unlike particle simulations, in hydrodynamic simulations  the
patterns tend to be very stable and regular for small as well as
for moderate amplitudes.  The growth of the selected wavelength
develops within 10 cycles after the perturbation has been imposed
 \citep[see][]{Movie}, 
and one observes only minor changes after this time. The periodicity of
the pattern is twice the shaking period, as corresponds to the
typical period doubling of this instability. Figure \ref{snaps}
shows the appearance of the pattern in a sequence of eight frames
which covers two periods of shaking and the corresponding frames
generated from the MD simulation.

The main difference between the HD and MD sequences in Figure \ref{snaps} is the
different landing time of the granular bed. The fourth and the eighth
snapshots are taken when the material visibly starts to deposit, which
happens at different times in HD and MD simulations. This deserves a careful
discussion. Indeed, one of the typical features accompanying the instability is
the gap formed between the material and the bottom wall, induced by an
acceleration of the plate at $\Gamma > 1$. In real experiments as well as in MD
simulations -- and leaving aside the effects of particle noise, there is a
true empty space opening and closing periodically. In HD simulations the
density is never very close to zero at the bottom wall. This fact has to be
attributed to the implemented boundary condition. More will be said in this
respect later in this section.  

\begin{figure}
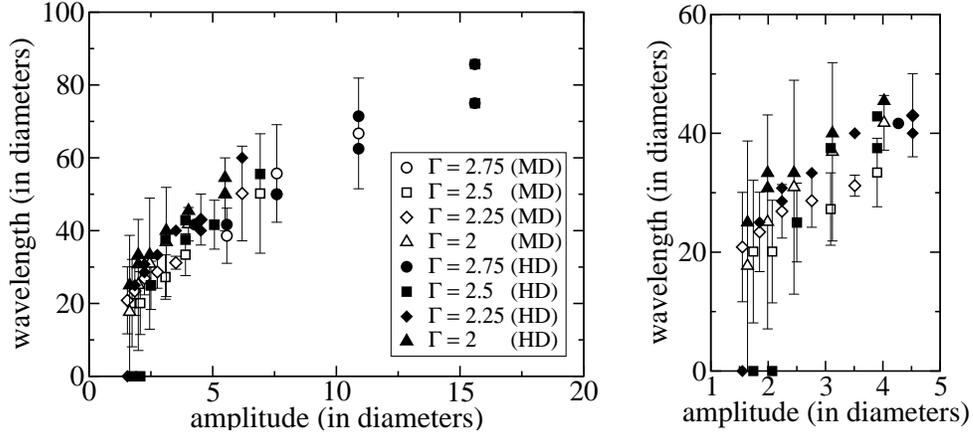

  \centerline{\includegraphics[bb=20 20 592 438,width=0.7\textwidth,clip]{FIGS/lambda-A.eps}\hspace{-1cm}\includegraphics[width=0.32\textwidth,clip]{FIGS/lambda-A.zoom.eps}}
\caption{The wavelength of the pattern as a function of the
amplitude of shaking, for different reduced accelerations
$\Gamma$. Open symbols: Numerical solution of the hydrodynamic
equations; full symbols: Event-driven molecular dynamics simulations. On the
right: A zoom of the  figure at low amplitudes. Error bars are drawn only for
MD data points.} 
\label{bifurcation}
\end{figure}

The bifurcation diagram is shown in Figure \ref{bifurcation}, from
which one can easily observe the extent of the agreement that both
types of simulations can provide, with results which
quantitatively diverge by no more than 20\%. Close to the
bifurcation point though, there is an important exception: According to
the hydrodynamic simulations, the pattern starts to form between
1.55 and 1.64 diameters of amplitude, since below $A = 1.64\sigma$ it is
not observed for any reduced acceleration $\Gamma=A(2\pi f)^2/g$ in the range
from 2.0 to 2.75, where $g$ is the acceleration of gravity.  For the MD
simulations, 
however, one obtains a few data points in the region of very low
amplitudes giving nonzero wavelengths. Nevertheless the errors
associated with these few MD measures are particularly large,
coming from the procedure to determine the characteristic pattern
wavelength in MD simulations: due to the general contribution of
particle noise, the wavelength was extracted from the histogram of
the Fourier analysis of the particle positions over 50 periods.
Otherwise, in the low amplitude regime, no pattern would be
guessed from the naked-eye inspection of the MD sequences.
Therefore, the typical histograms contain a wide spread of
wavelengths in this regime, which reflects in the standard
deviations ($d\lambda$ in Table \ref{tab:parameters}) plotted in
Figure \ref{bifurcation} as error bars. 
 The duplicate data points in some of the
hydrodynamic samples refer to situations where  a child peak develops or
disappears (due to finite size effects introduced by the periodic boundaries)
and as a result, both wavelengths can be observed in the course of one
simulation. This tends to happen  in the large amplitude regime, where the
pattern is less stable.

\begin{figure}
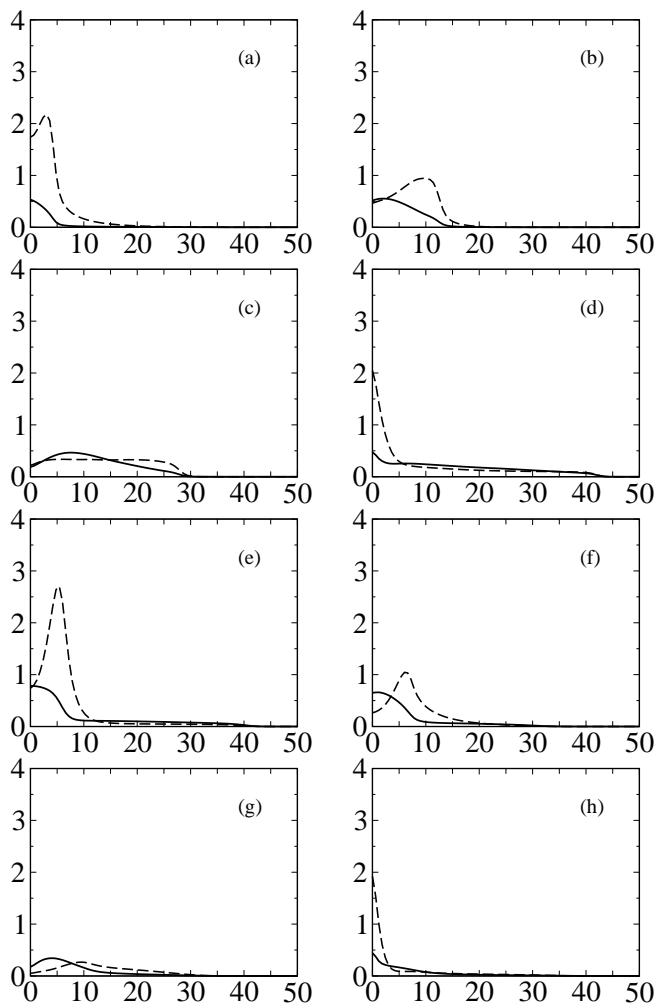

  \centerline{
    \includegraphics[bb= 52 24 482 373,width=0.30\textwidth,clip]{FIGS/T21HD.eps}
    \hspace{8pt}
    \includegraphics[bb= 52 24 482 373,width=0.30\textwidth,clip]{FIGS/T34HD.eps}}
  \centerline{
    \includegraphics[bb= 52 24 482 373,width=0.30\textwidth,clip]{FIGS/T46HD.eps}
    \hspace{8pt}
    \includegraphics[bb= 52 24 482 373,width=0.30\textwidth,clip]{FIGS/T59HD.eps}}
  \centerline{
    \includegraphics[bb= 52 24 482 373,width=0.30\textwidth,clip]{FIGS/T71HD.eps}
    \hspace{8pt}
    \includegraphics[bb= 52 24 482 373,width=0.30\textwidth,clip]{FIGS/T84HD.eps}}
  \centerline{
    \includegraphics[bb= 52 24 482 373,width=0.30\textwidth,clip]{FIGS/T96HD.eps}
    \hspace{8pt}
    \includegraphics[bb= 52 24 482 373,width=0.30\textwidth,clip]{FIGS/T09HD.eps}}
  \caption{Hydrodynamic simulation: A 2-period sequence of the vertical
    profiles of the packing fraction $\nu$ (solid line) and the rescaled
    internal energy $\rho T/\sigma g$ (dashed) as a function of height, in
    diameters, for the same times as the snapshots in Figure \ref{snaps}(left). }
    \label{dens-tempHD}
\end{figure}

\begin{figure}
\centerline{\includegraphics[width=0.60\textwidth,clip]{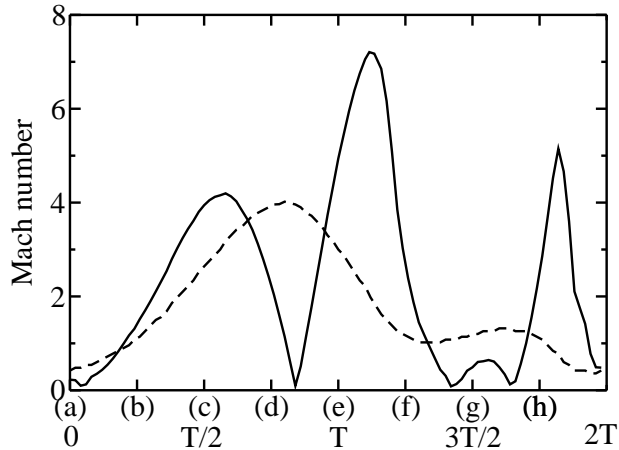}}
\caption{Solid line: Evolution of the Mach number over two periods for the
  frames in   Figure \ref{dens-tempHD} at a height (variable) where the
  packing fraction is 5\%. Dashed line: height (in diameters divided by 10)
  corresponding to the values of the Mach shown.}
\label{mach}
\end{figure}

In Figure \ref{dens-tempHD} we show a 2-period sequence of the
vertical profiles of the packing fraction $\nu$ and the rescaled
internal energy $\rho T/\sigma g$, as a function of height in
diameters. For the plots we have selected one of the locations in
Figure \ref{snaps} where a valley/cusp develops alternatively. The
first snapshot (a) shows the granular material already being
compressed by the incoming bottom wall. As one can appreciate, the
extent of the layer (solid line) is small so the profile
corresponds to the location of a valley at this time. The thermal
energy is high and has started to propagate upwards after the
impact. In Figure \ref{dens-tempHD}(b) the layer is thickening while
at the peaks is thinning, as (f) shows. In (c) the energy has
already propagated to the top, the material is suspended and
rather fluidized, and the cusp is visible. In (d)  another impact
is taking place, the density is growing at the bottom, and one
more shock wave has been generated whose energy extincts as the
peak redissolves to feed the closest locations where new cusps are
forming -- indeed, in (e) the material is close packed at the bottom, and thus
density can only further decrease by pressure gradients. From the comparison
between (c) and (g), one can see that the densities are very different in
peaks and valleys, and the shock progresses  thus differently. The careful
analysis of the shock wave propagating through a (homogeneous) vibrated bed
has been done by \cite{BMSS02} and here suffices to point at the fact that the
flow is indeed supersonic, with Mach numbers in the range from 0 to 10. See
the Figure \ref{mach} to observe the typical oscillations of the Mach
number at a valley / cusp. The highest values of the Mach number are achieved
 in the dilute phase but not far away from the clustered phase; the figure
 shows the Mach number recorded at the height where the packing fraction is 5\%, as a
 convenient measure of the  interface between the dense and rarefied
 material.  The convective motion along the $x_1$ direction which removes
 material from the dense towards the dilute locations will be analyzed later.

\begin{figure}
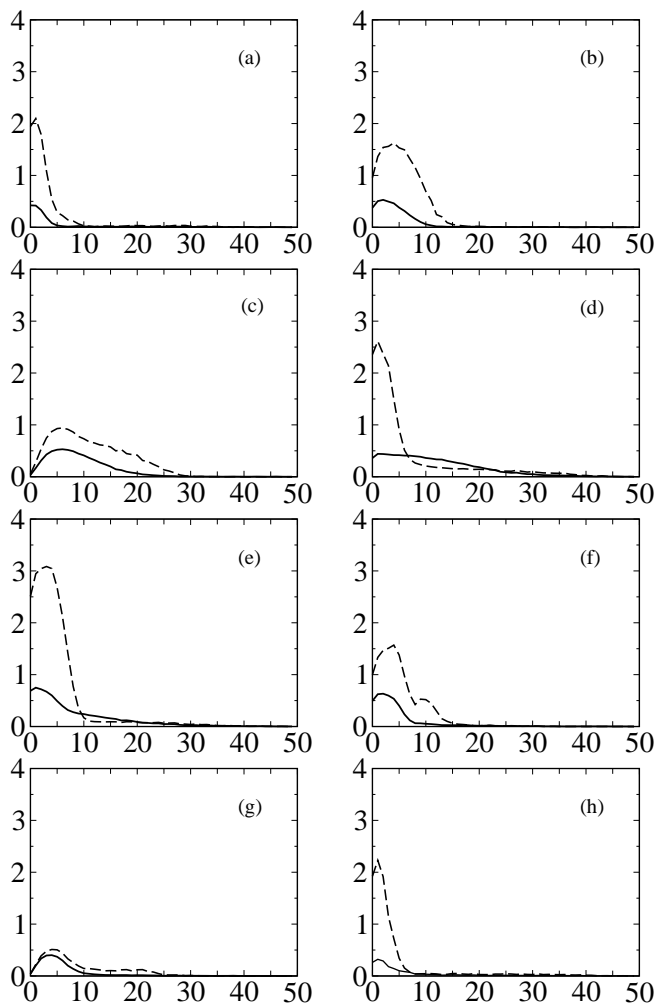

\centerline{
  \includegraphics[bb= 52 24 482 373,width=0.30\textwidth,clip]{FIGS/T96MD.eps}
  \hspace{8pt}
  \includegraphics[bb= 52 24 482 373,width=0.30\textwidth,clip]{FIGS/T09MD.eps}}
\centerline{
  \includegraphics[bb= 52 24 482 373,width=0.30\textwidth,clip]{FIGS/T21MD.eps}
  \hspace{8pt}
  \includegraphics[bb= 52 24 482 373,width=0.30\textwidth,clip]{FIGS/T39MD.eps}}
\centerline{
  \includegraphics[bb= 52 24 482 373,width=0.30\textwidth,clip]{FIGS/T46MD.eps}
  \hspace{8pt}
  \includegraphics[bb= 52 24 482 373,width=0.30\textwidth,clip]{FIGS/T59MD.eps}}
\centerline{
  \includegraphics[bb= 52 24 482 373,width=0.30\textwidth,clip]{FIGS/T71MD.eps}
  \hspace{8pt}
  \includegraphics[bb= 52 24 482 373,width=0.30\textwidth,clip]{FIGS/T89MD.eps}}
\caption{Molecular Dynamics simulation: A 2-period sequence of the vertical
profiles of the packing fraction $\nu$ (solid line) and the
rescaled internal energy $\rho T/\sigma g$ (dashed) as a
function of height, in diameters, for the same times as the
snapshots in Figure \ref{snaps}(right). } \label{dens-tempMD}
\end{figure}

For the MD simulation, the time-dependent thermal energy has been
extracted by averaging over 50 cycles the corresponding phases of
the local fluctuational kinetic energy, as defined by Eq.
\eqref{temperature}. Afterwards an average over equivalent locations has been
performed.  The profiles are shown in 
Figure \ref{dens-tempMD}. Comparison between Figs. \ref{dens-tempHD} and
\ref{dens-tempMD} indicates that the quantitative difference in the
temperature profiles is small. However, the hydrodynamic profiles show a
sharper shock wave, as compared with their corresponding MD counterparts. This
feature is related to the residual heating observed in MD simulations at
heights where the granular layer is sufficiently dilute (very visible for
example in Figure \ref{dens-tempMD}f). Movies showing  the complete sequences are available as  \citep{Movie}. From their inspection one can realize 
that the maximum values of the temperature do not  differ very significantly,
whereas the MD profiles appear systematically thicker. This   
  phenomenon might be attributed in part to the  mesoscopic averaging procedure
employed here.  It is certainly known, on the other hand, that hydrodynamic
variables can be scale dependent in granular 
systems \citep{LNP564} due to the lack of scale separation. So even if
the number of samples (cycles) could  in principle be increased, and therefore
the mean fields better defined, the local noise will persist at scales of
the order of one particle (which is the grid size used here for averaging the
MD sequences), particularly in dilute regions. The number of samples is practically
limited by the fact that the MD pattern is not strictly stationary for very long times. It is not desirable either to
increase the grid size for averaging, for obvious reasons: Here,
where the pattern is of the size of a few grains, the fields obtained from MD
simulations would lose structure and definition. Oppositely, the granular
Navier-Stokes equations \eqref{ns} do not account for such a source of
fluctuations. However, and even with these restrictions, one can conclude that
good quantitative agreement can be achieved with mesoscopic statistics: a
manageable number of samples (25) and a one-diameter characteristic cell size,
 along with a final average of means over corresponding locations. 

There is still another difference between the MD and HD profiles,
and it is related to the gap formed when the granular layer takes
off (Figure \ref{snaps}c and g). In MD simulations, during  the
stage of downward displacement of the wall, the entire layer is
levitating, as one can observe from the corresponding frames in Fig.
\ref{dens-tempMD} by noting that the density at zero
height is exactly zero. However, in HD simulations the density at
the bottom is never very close to zero. It is particularly large
at the base of the cusps, (Figure \ref{dens-tempHD}c), where the
packing fraction is  0.19. Seemingly, the hydrodynamic layer is
more sticky, or behaves more inelastically, than the MD granular
bed. Besides that, in hydrodynamic simulations, the fact that there
is some  material stuck to the bottom during the take off anticipates
the impact with the wall and the corresponding generation of the shock
wave. This is the source of the mismatch between the landing times, to which
 Figure \ref{snaps} refers.

\begin{figure}
  \centerline{
    \includegraphics[width=0.40\textwidth,clip]{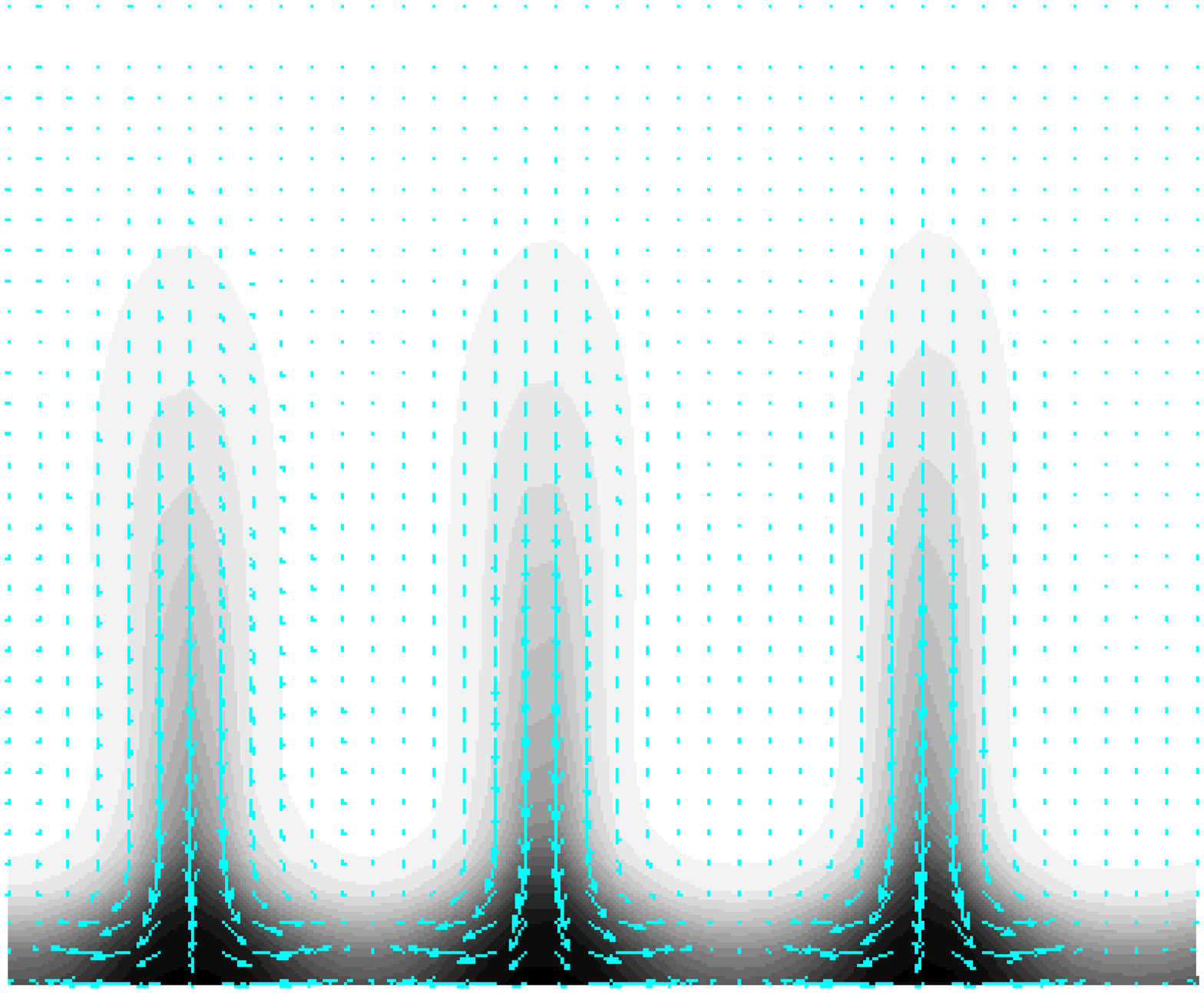}
    \hspace{8pt}
    \includegraphics[width=0.40\textwidth,clip]{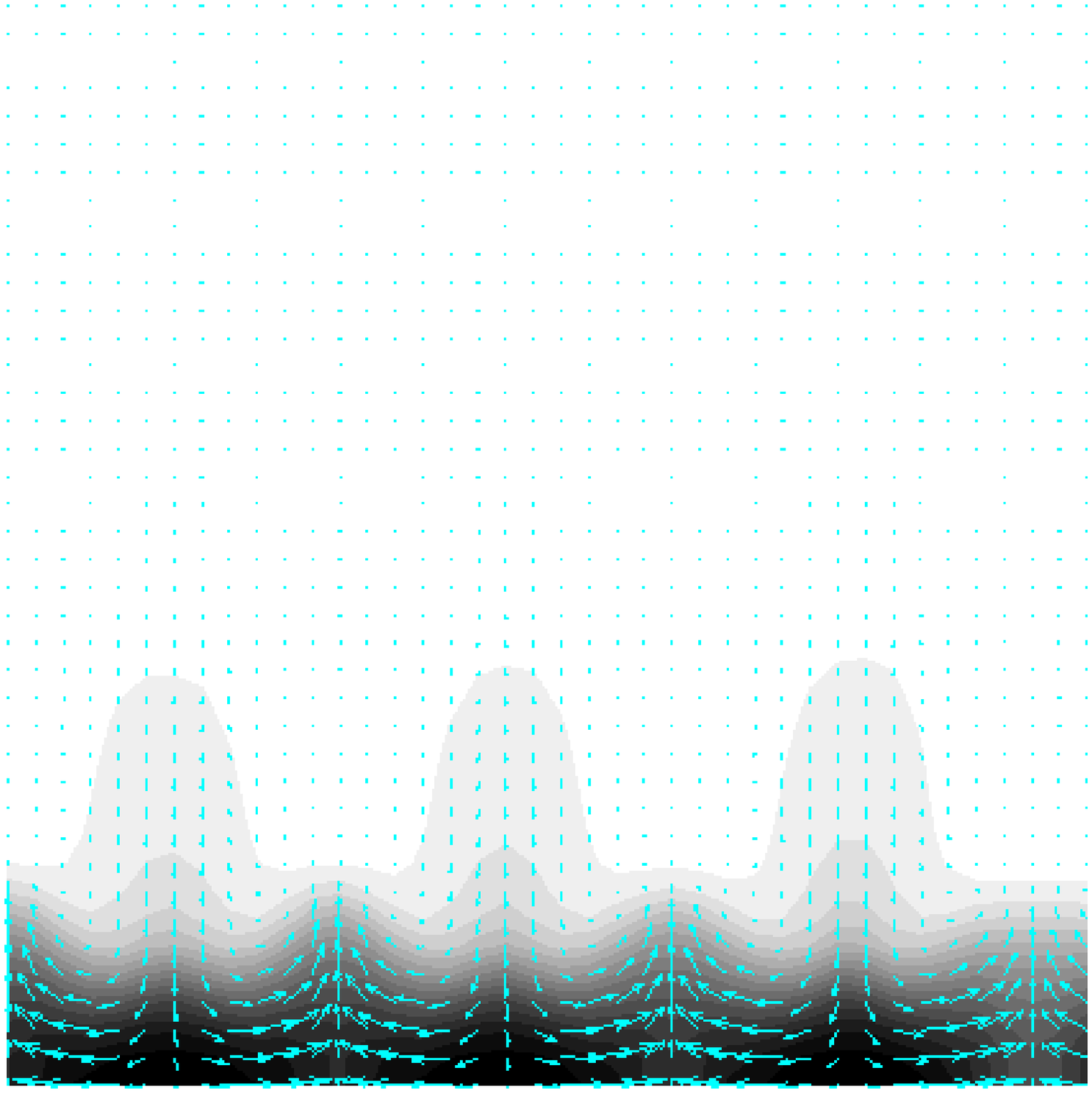}}
  \centerline{
    \includegraphics[width=0.40\textwidth,clip]{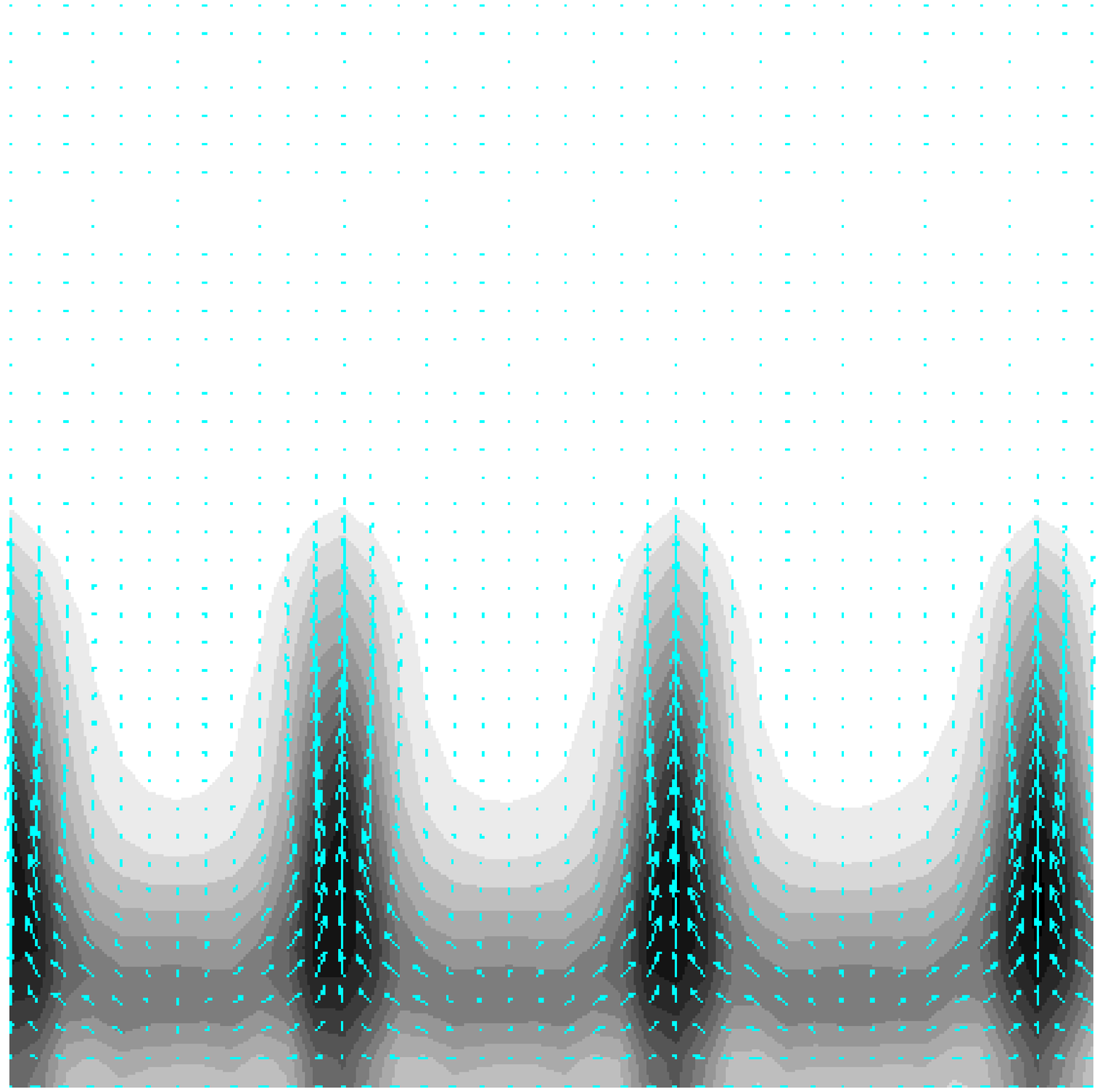}
    \hspace{8pt}
    \includegraphics[width=0.40\textwidth,clip]{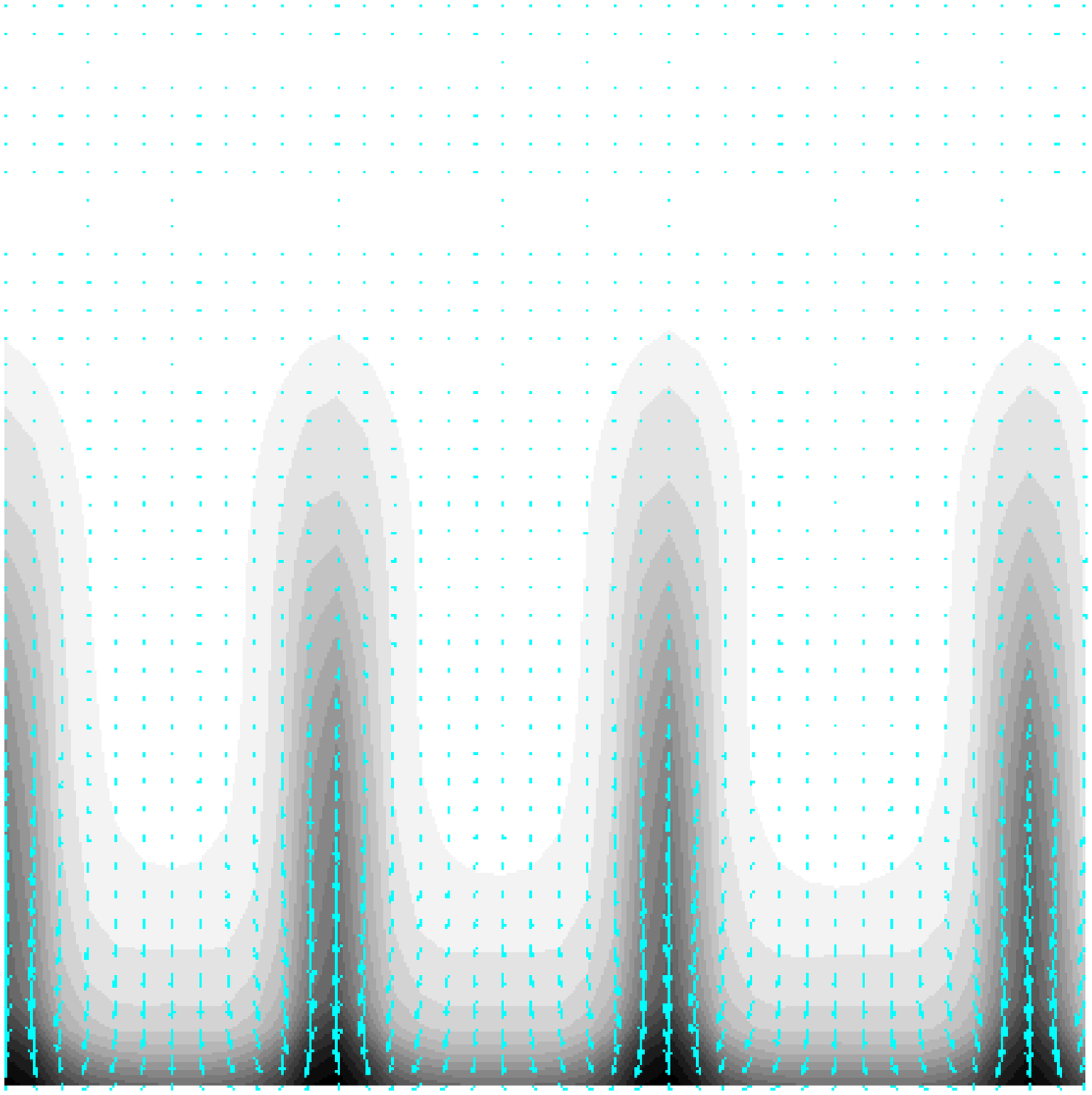}}
  \centerline{
    \includegraphics[width=0.40\textwidth,clip]{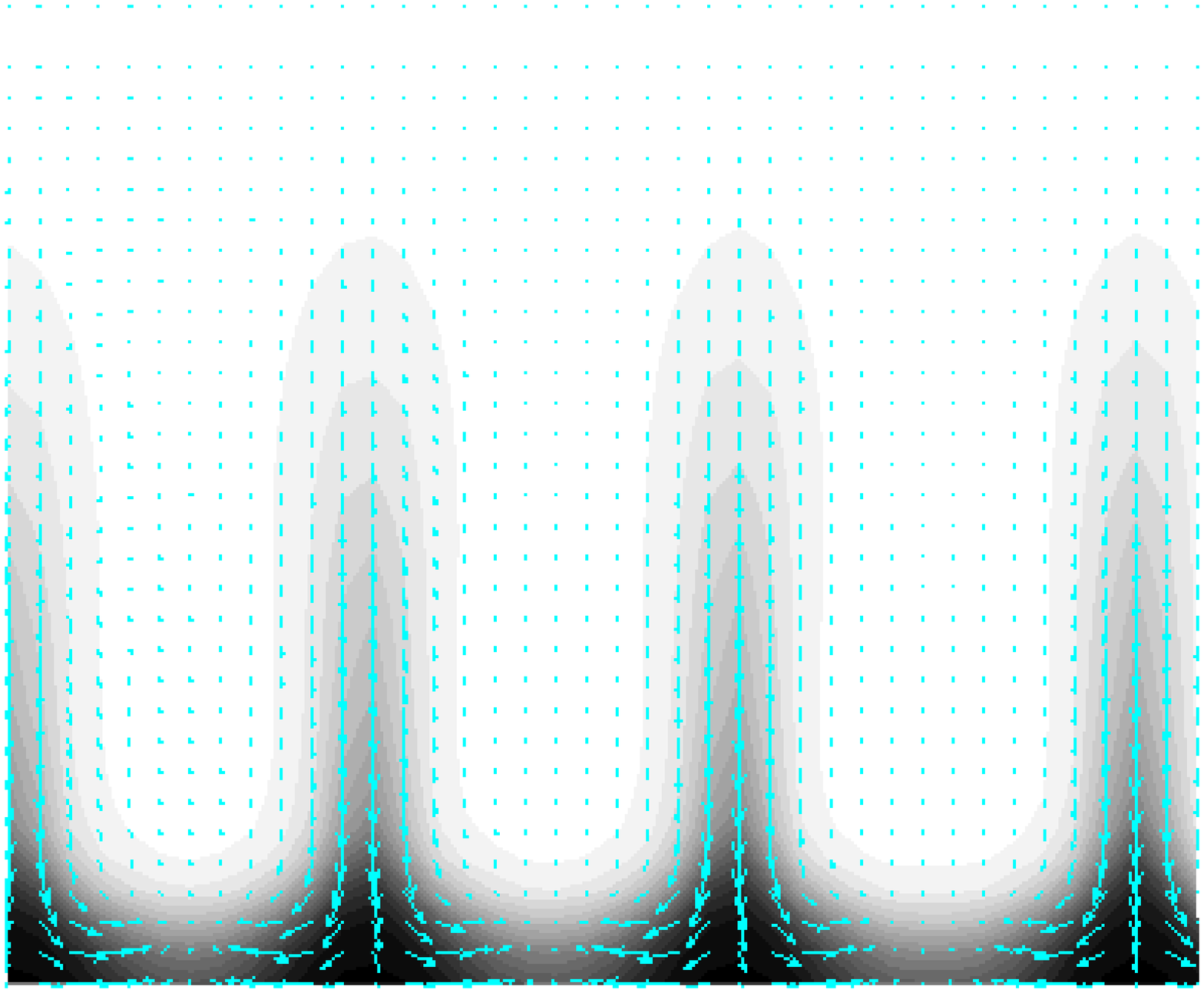}
    \hspace{8pt}
    \includegraphics[width=0.40\textwidth,clip]{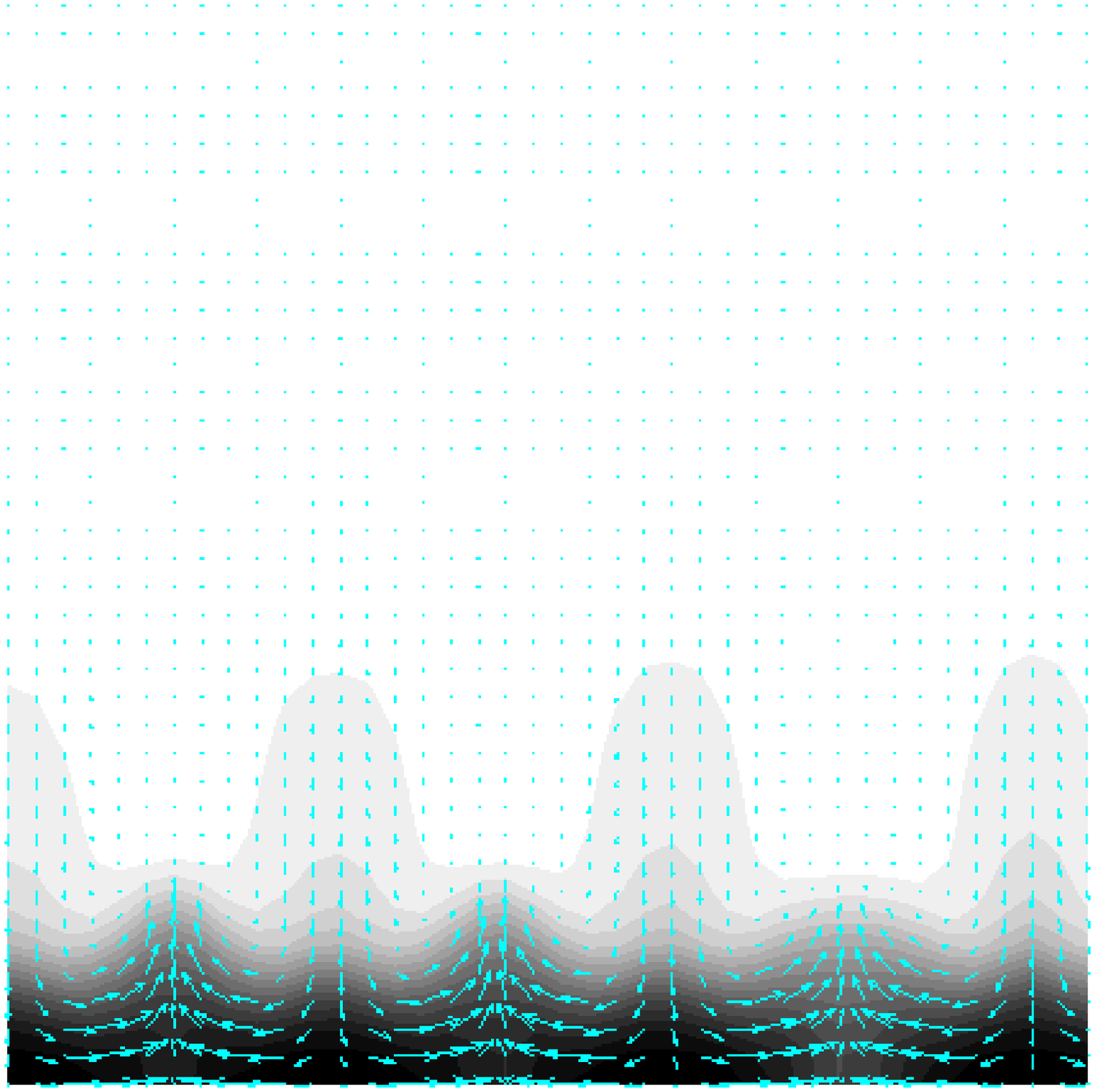}}
  \centerline{
    \includegraphics[width=0.40\textwidth,clip]{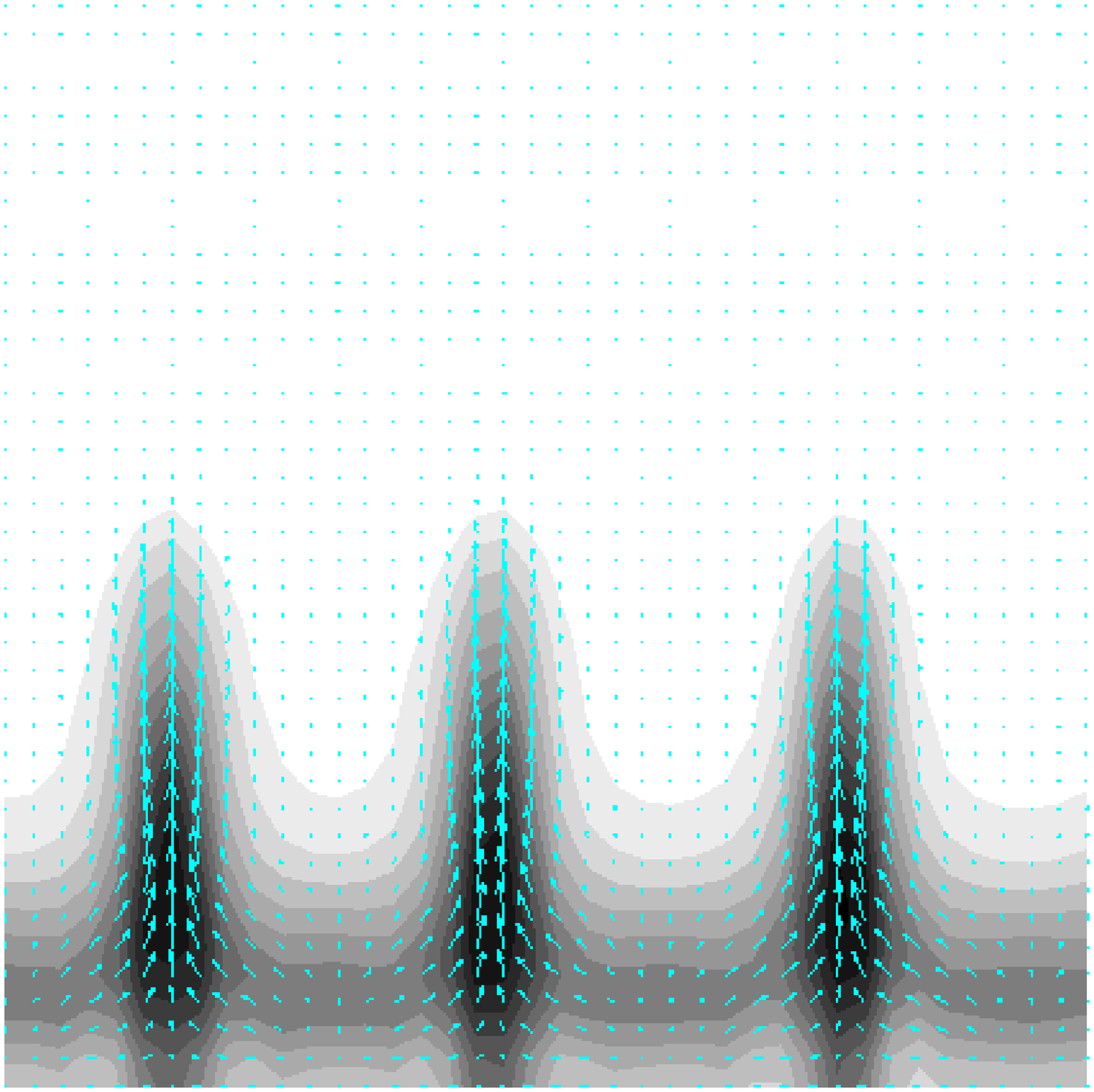}
    \hspace{8pt}
    \includegraphics[width=0.40\textwidth,clip]{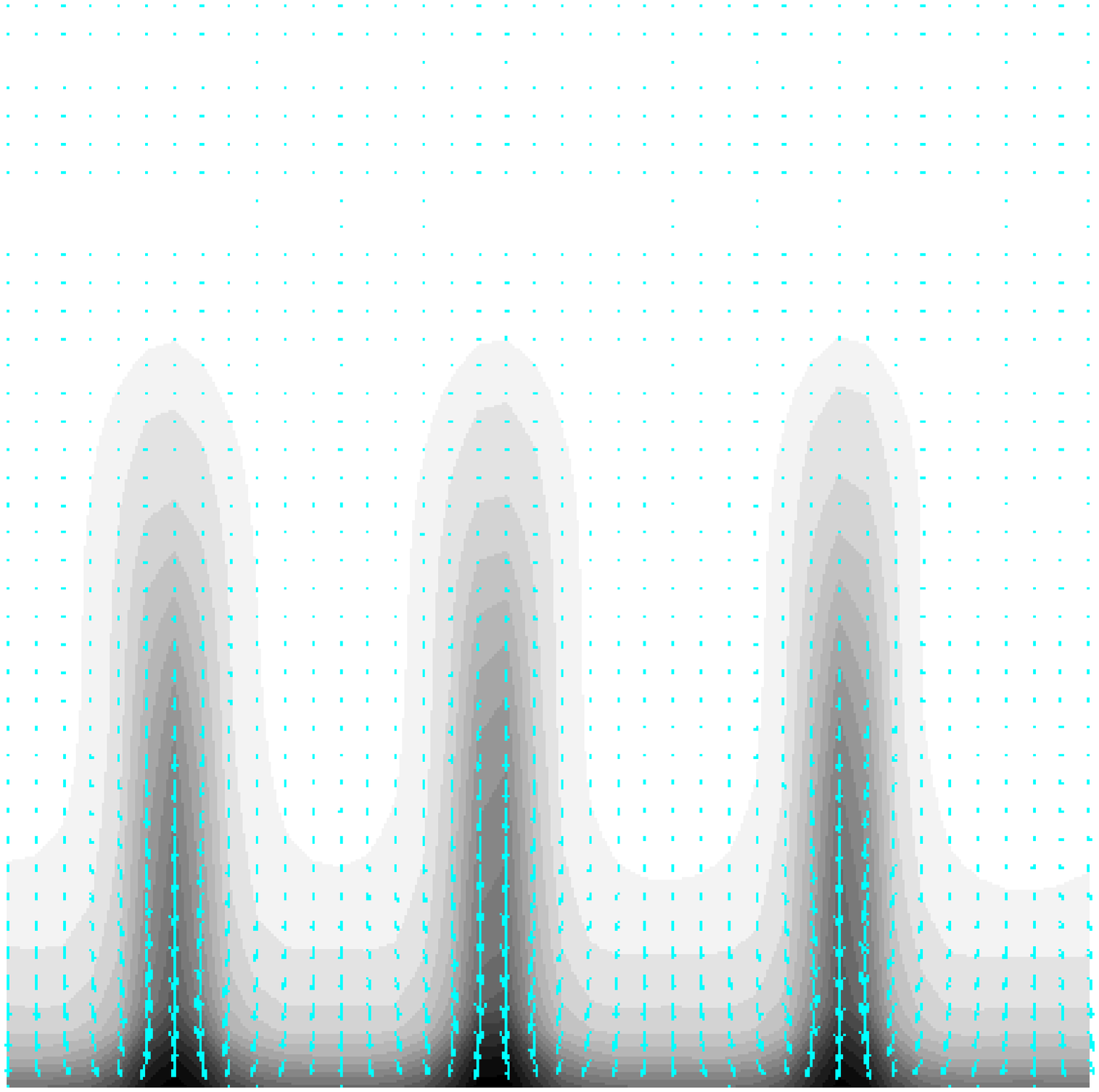}}
  \caption{Hydrodynamic simulation: The linear momentum plotted over the
    density field in gray shades. The frames are taken at the same times  as
    in Figure \ref{snaps}(left) and cover two cycles of the driving
    oscillation. The complete sequence is available as supplementary material.} 
  \label{vHD}
\end{figure}

\begin{figure}
  \centerline{
    \includegraphics[width=0.40\textwidth,clip]{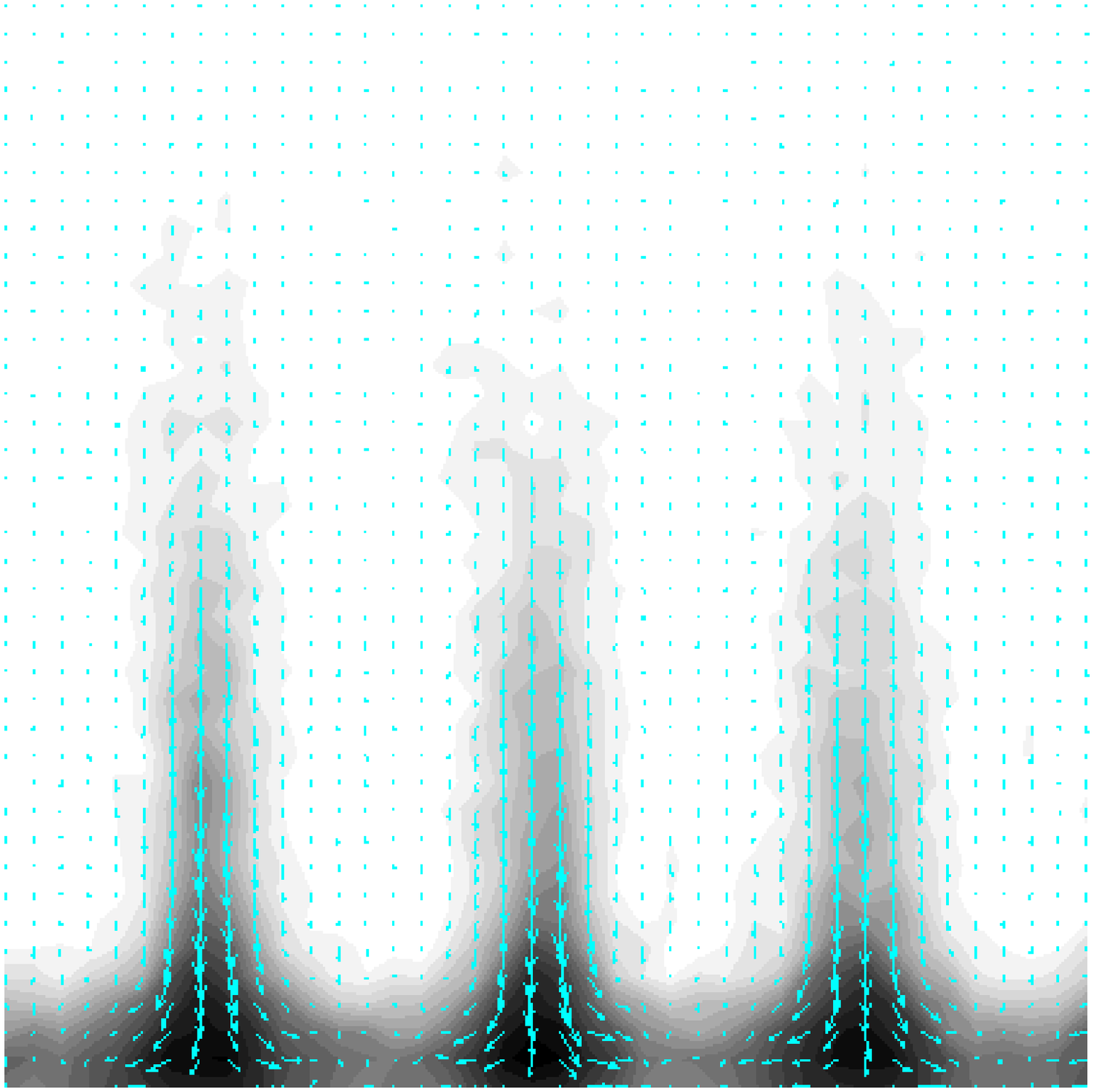}
    \hspace{8pt}
    \includegraphics[width=0.40\textwidth,clip]{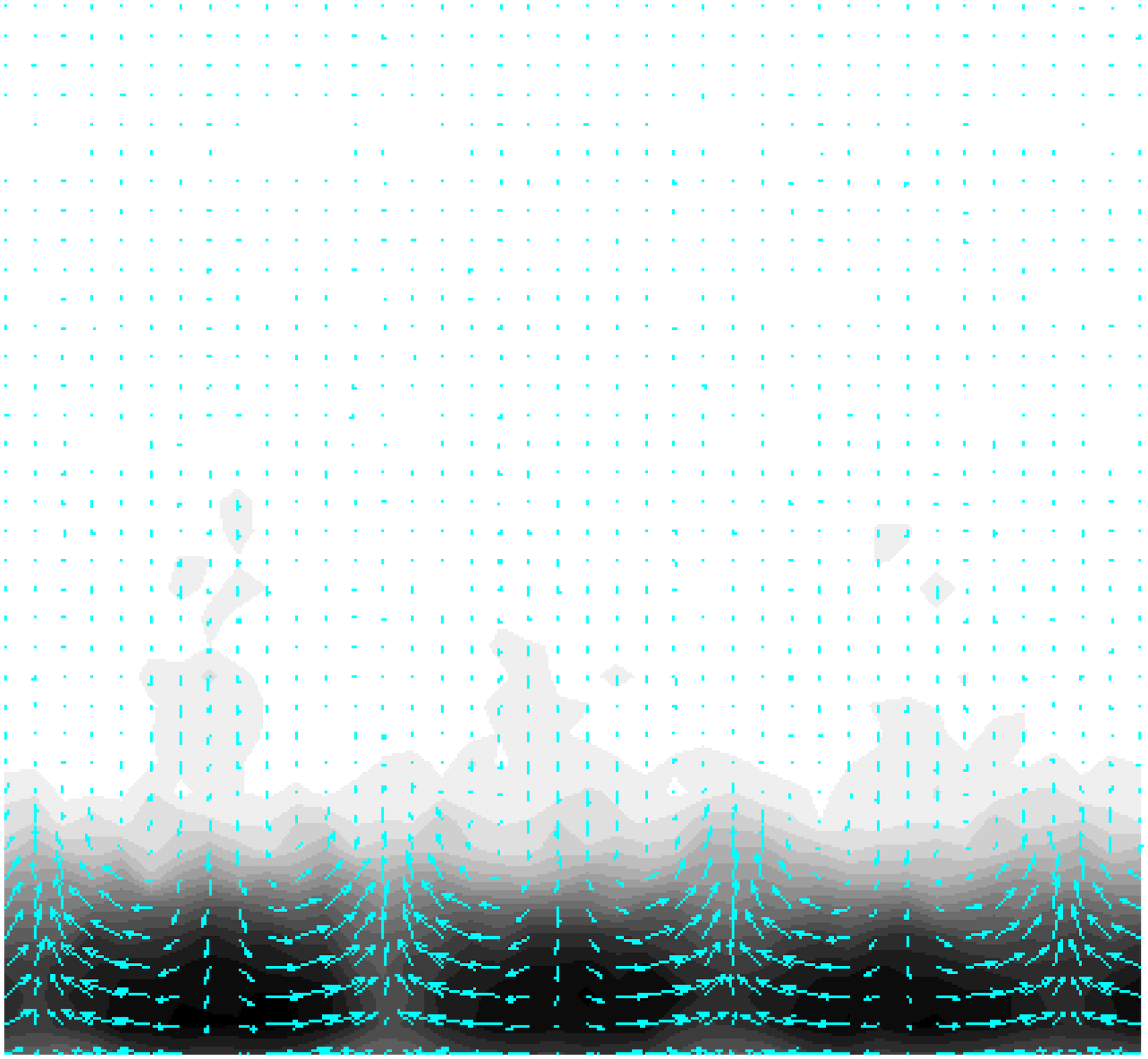}}
  \centerline{
    \includegraphics[width=0.40\textwidth,clip]{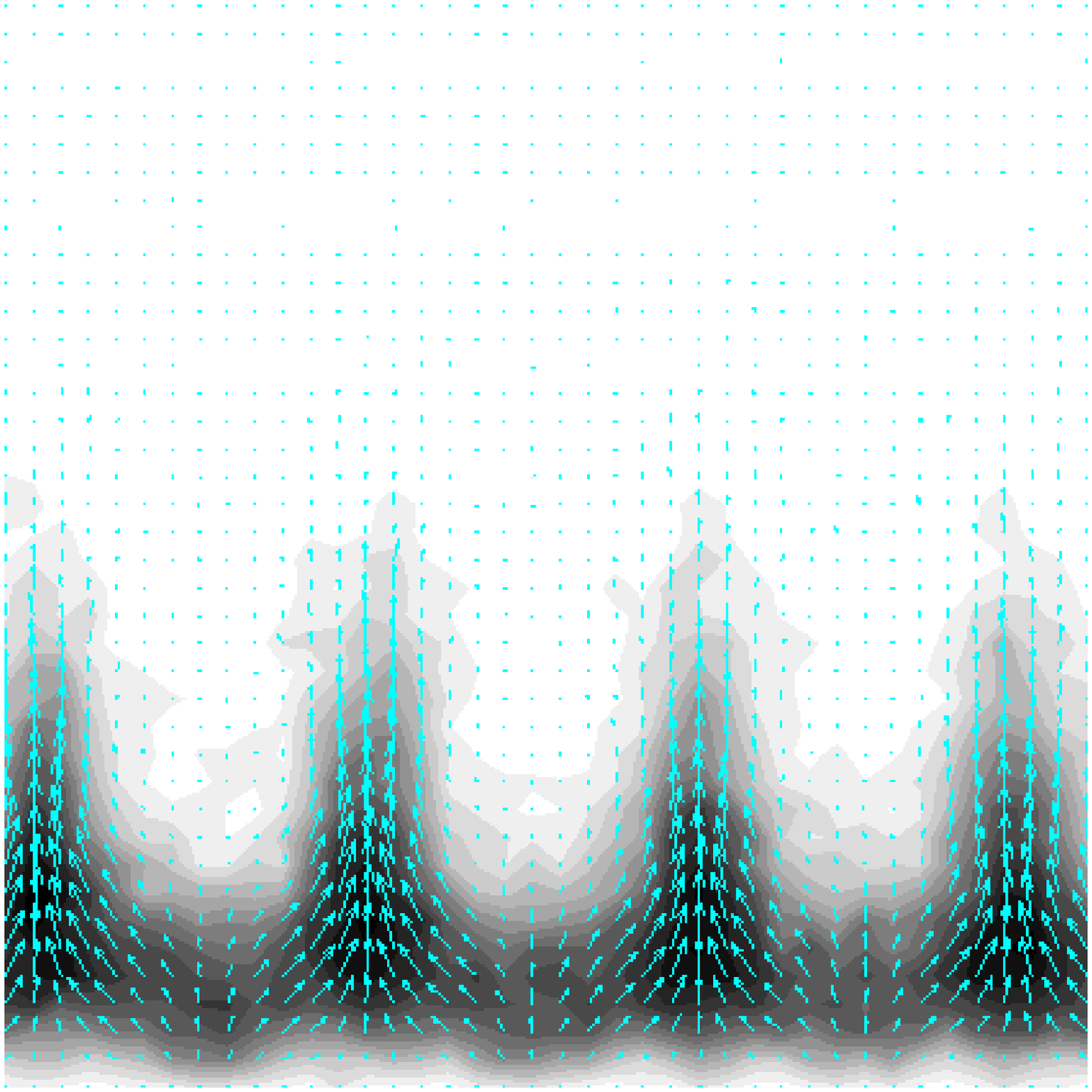}
    \hspace{8pt}
    \includegraphics[width=0.40\textwidth,clip]{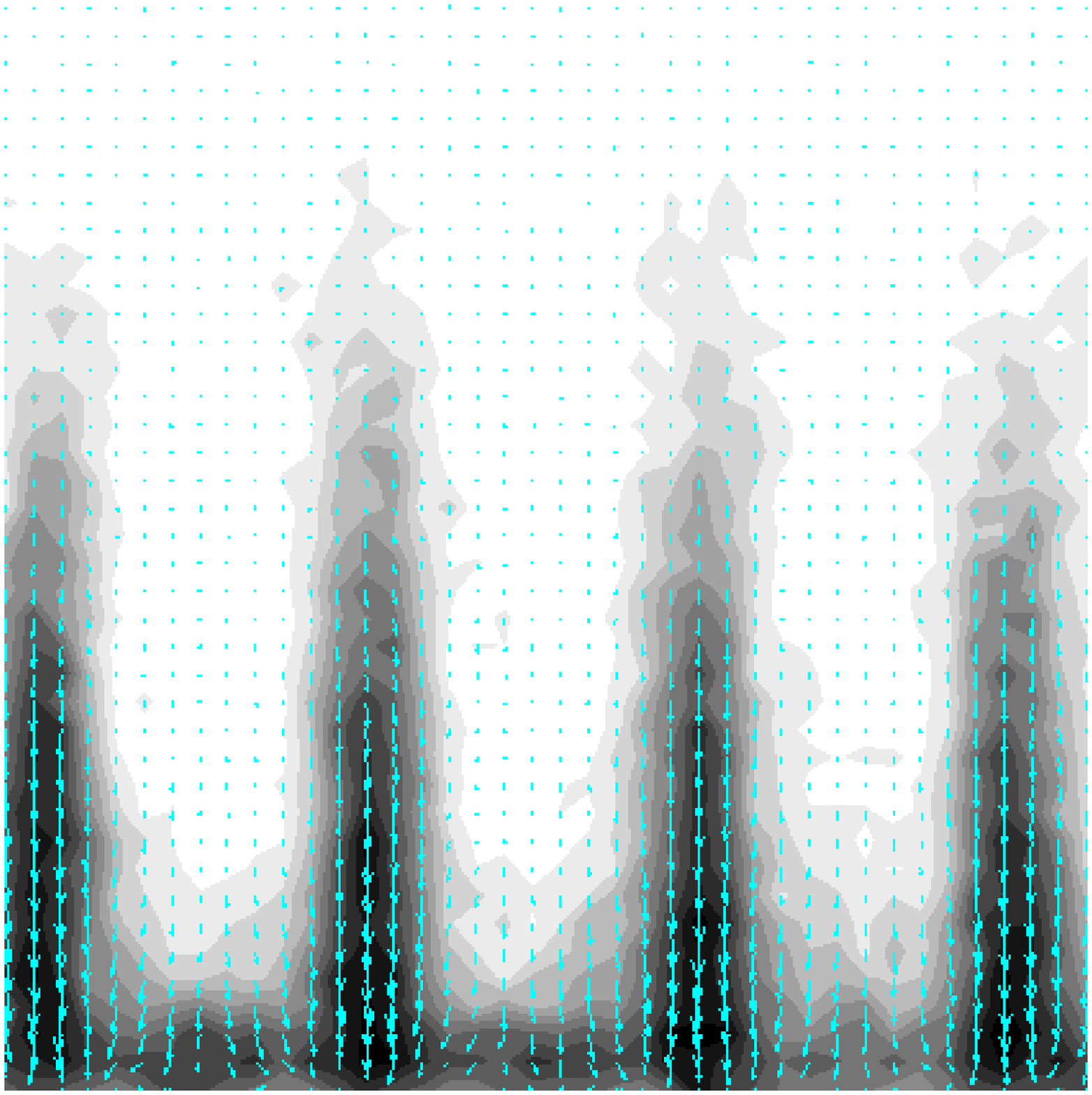}}
  \centerline{
    \includegraphics[width=0.40\textwidth,clip]{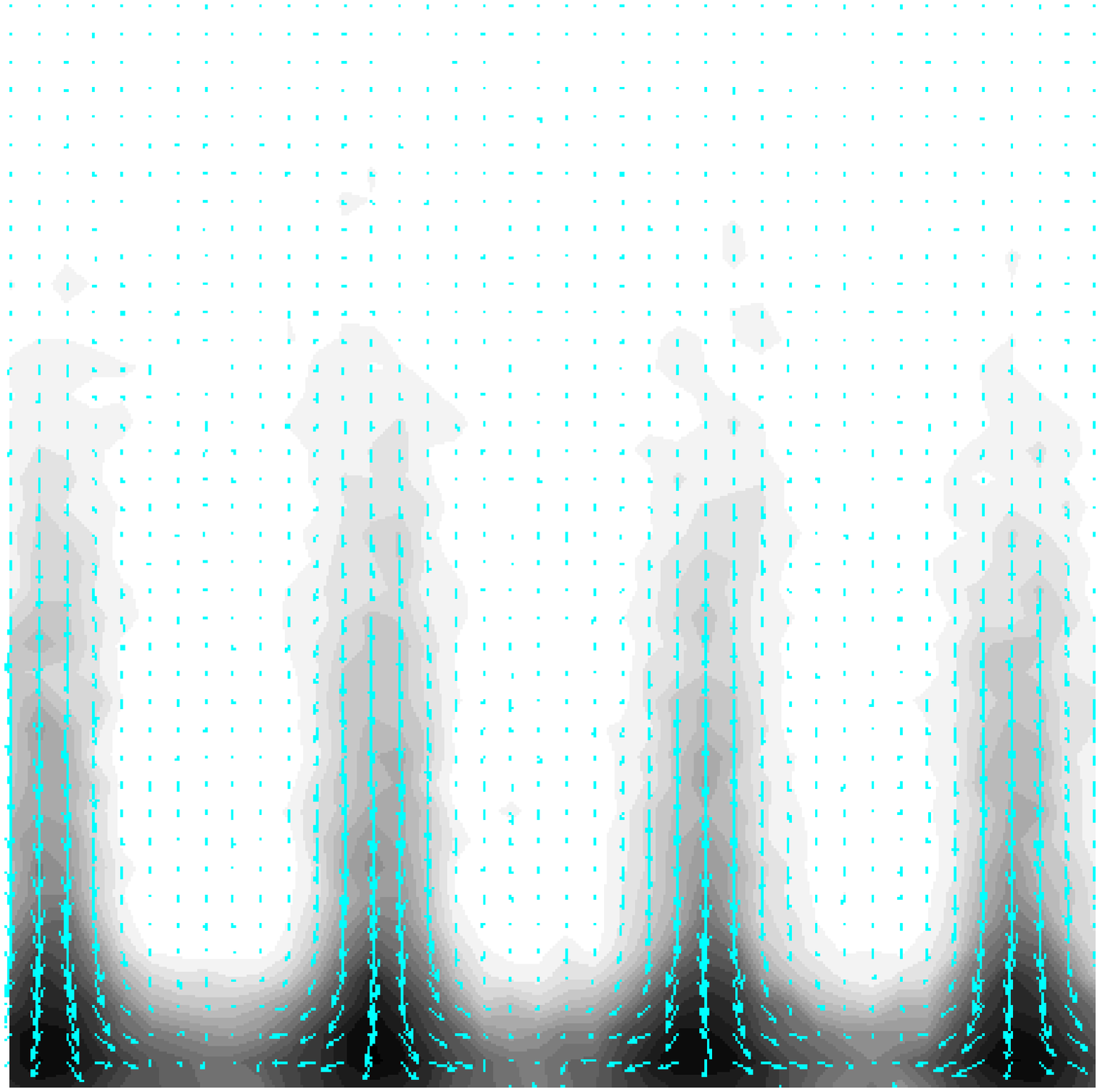}
    \hspace{8pt}
    \includegraphics[width=0.40\textwidth,clip]{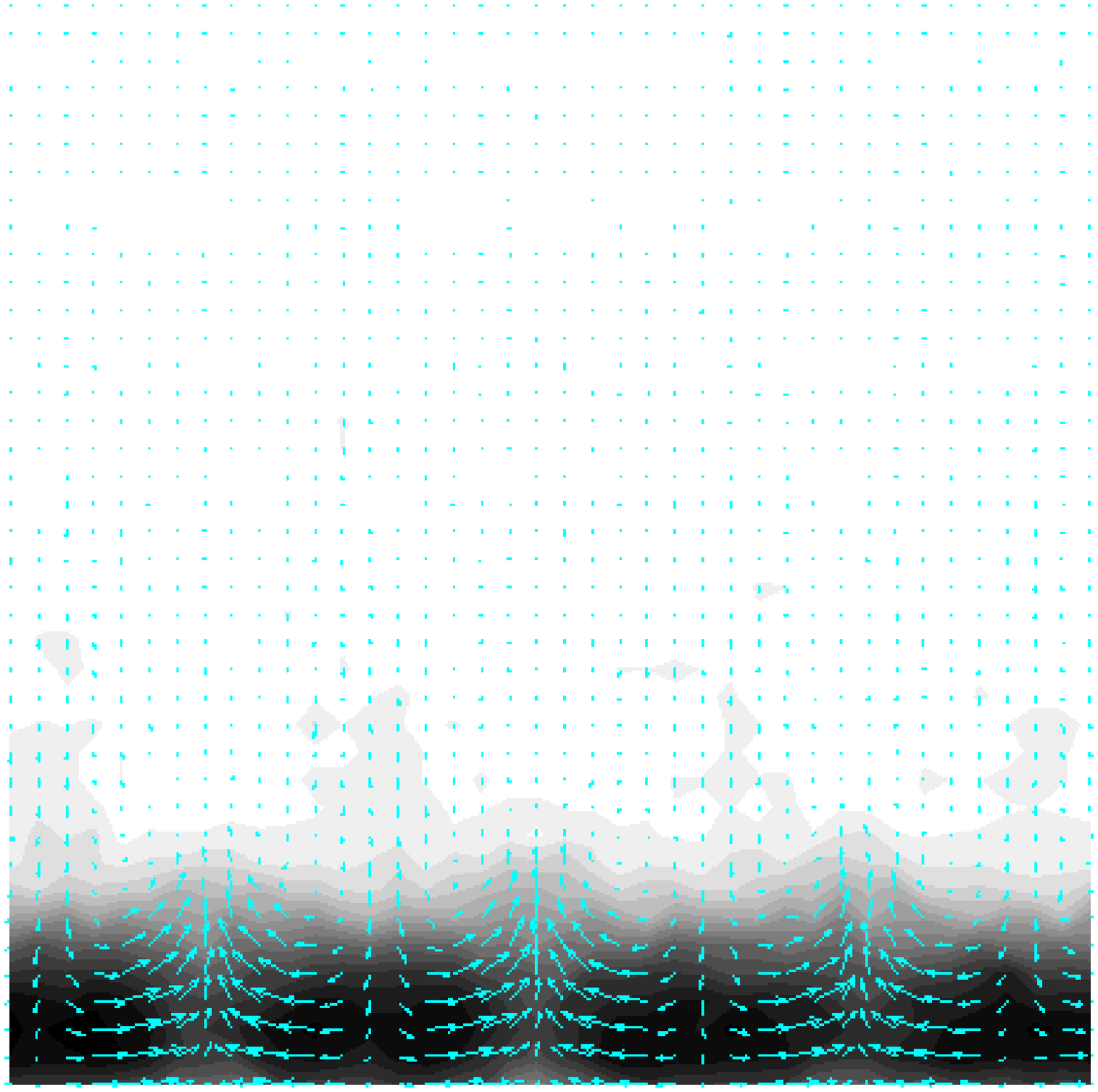}}
  \centerline{
    \includegraphics[width=0.40\textwidth,clip]{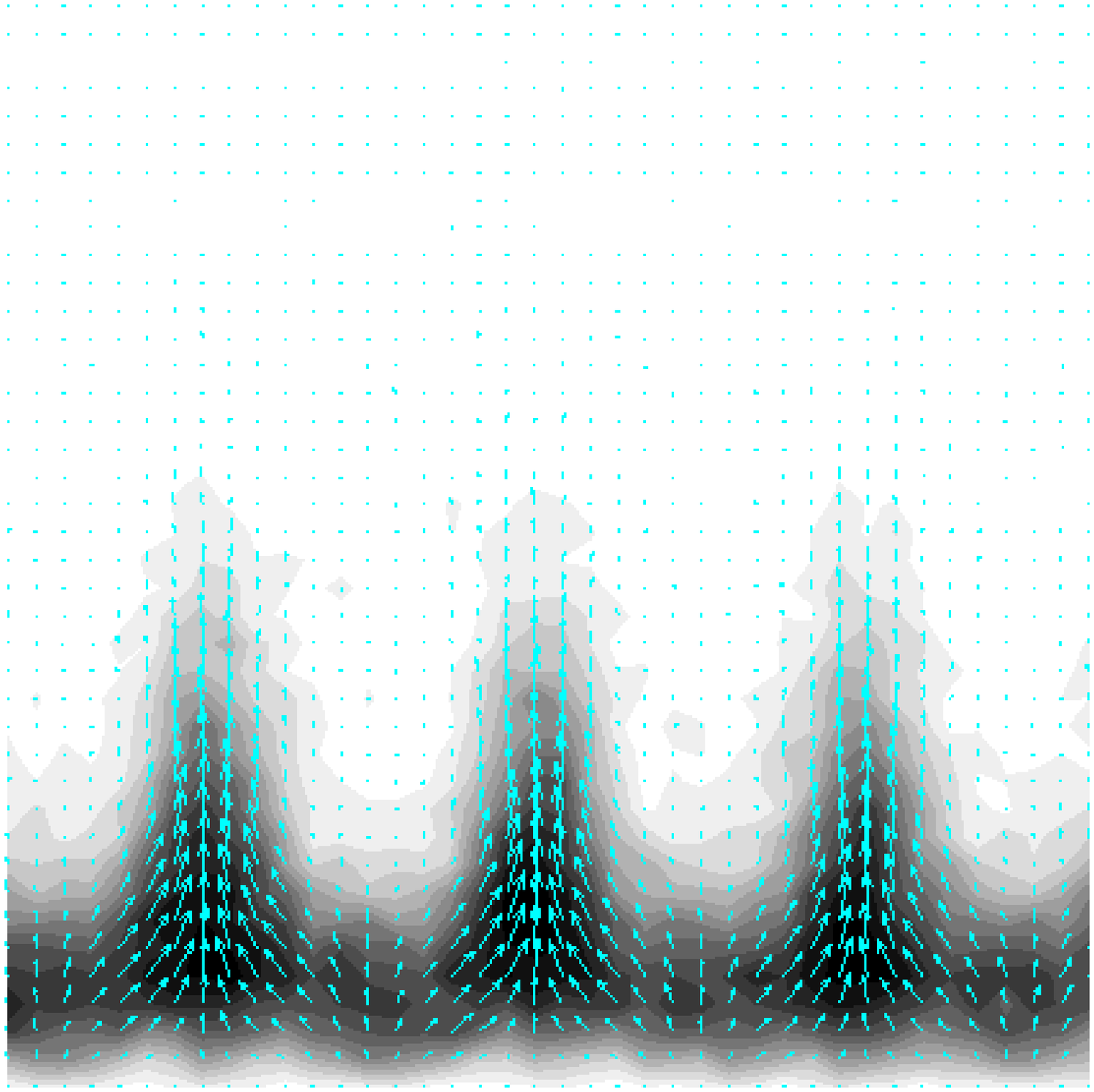}
    \hspace{8pt}
    \includegraphics[width=0.40\textwidth,clip]{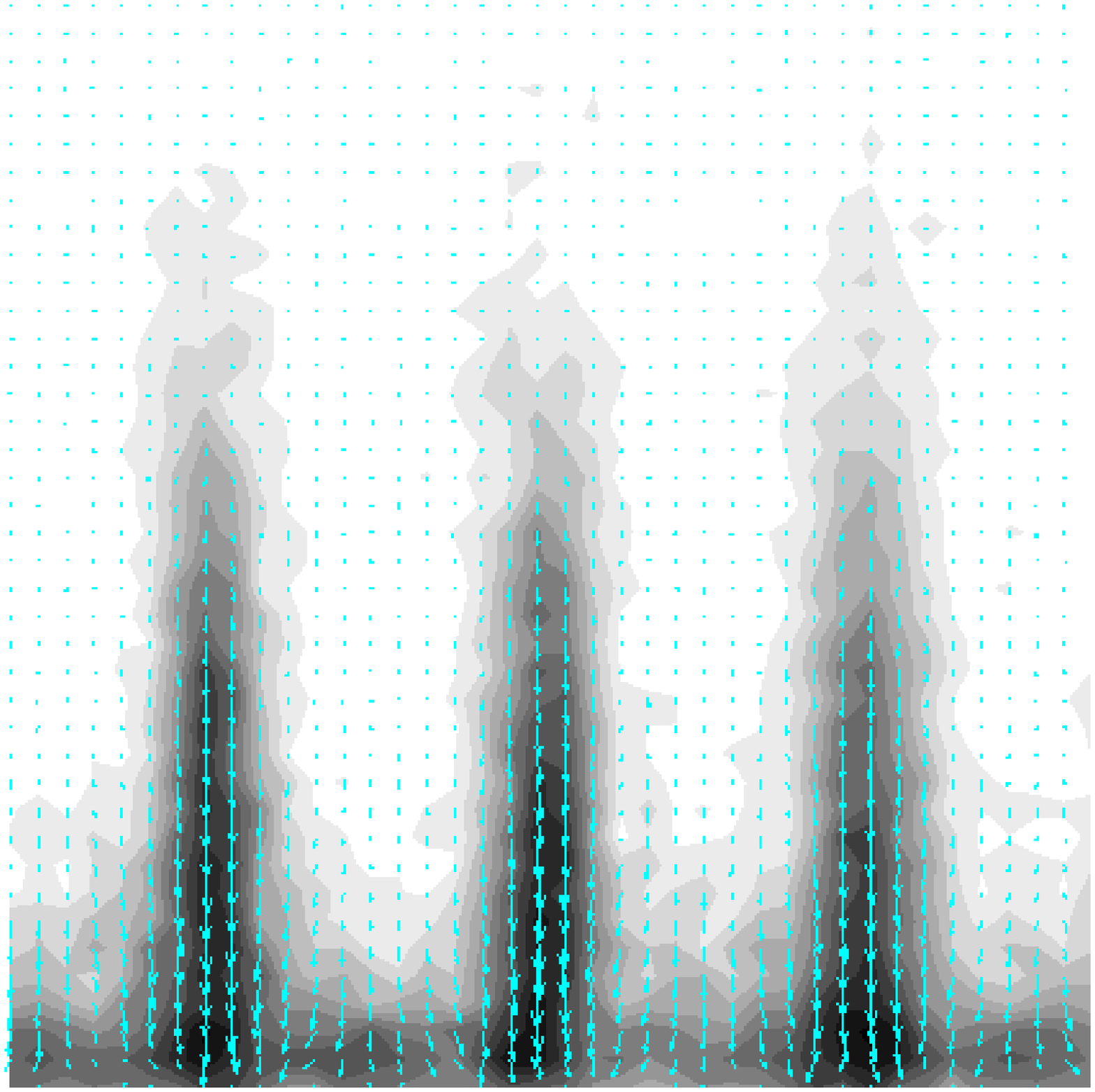}}
  \caption{Molecular Dynamics simulation: The linear momentum plotted over the
    density field in gray shades. The frames are taken at the same times in
    Figure \ref{snaps}(right) and cover two cycles of the driving oscillation. }
  \label{vMD}
\end{figure}

It is worth to analyze the velocity field and the convection
pattern which accompanies the Faraday waves. These are shown in
values relative to the velocity of the plate in Figs. \ref{vHD}
and \ref{vMD} for hydrodynamic  and MD simulations, respectively, and in the \citeauthor{Movie}.
In fact what is shown is the linear momentum field rather than the
velocity, in order to focus onto the most relevant part of the
system eliminating from the picture the almost empty regions.
Comparing both panels we see again the effect that was pointed
at above: in hydrodynamic simulations, the material never abandons
contact with the vibrating plate. This feature has been observed
before in hydrodynamic  simulations of vibrated granular layers without
pattern formation 
\citep{Shattuck}. Besides that, in Figs.  \ref{vHD} and \ref{vMD}
we observe the removal of material from the dense towards the
loose regions after each impact of the wall, during the alternate
formation of cusps. The scaling factor  for the arrows -- chosen
identical for both figures, again reveal good quantitative agreement of the hydrodynamic simulations and MD. In HD we
obtain slightly lower values of the linear momentum as compared with MD
simulations, by a factor which is shown to be about 20\% -- the maximum value
achieved by the linear momentum was 1.17 in some units for the MD sequence;
for the HD simulation was 0.935.  We recall that simple reflecting boundaries
are used for the hydrodynamic  simulations (with zero no-slip), and that
particles are perfectly smooth in  MD simulations. It is difficult to evaluate
the effect of the choice of boundary conditions in more complex setups, but in
view of the results, it seems clear that this undisturbed sliding along the 
plate looks like the only  reasonable choice in this very simple
situation. 

\section{Conclusions}

Up to now, we have compared the results given by the two
simulation methods considered: The traditional approach by
event-driven MD and the hydrodynamic approach by means of a CFD
code. We have shown that qualitatively and quantitatively as well,
adequately armored  hydrodynamic codes can reproduce the features
of particle simulations in complex flow problems dealing with
transient strong shock waves and pattern formation, such as the
Faraday instability here investigated. In particular, the
appearance and wavelength of the pattern has been reproduced with
a very good approximation as seen from the bifurcation diagram
(Figure \ref{bifurcation}). The hydrodynamic fields compare
very well to the corresponding  averaged particle fields (with the
exception of the  gap formed during the take off of the granular
layer). The hydrodynamic temperature is shown to agree with its MD counterpart
 to which ensemble averaging at a typical grid size of one particle diameter
 has been applied. Last, the
velocity field obtained in hydrodynamic simulations follows the
periodic alternancy of the pattern and shares the type of
structure observed in the particle velocity field, if however
giving somewhat smaller values. The discrepancies found along the analysis
are of the order of 20\%.

It is worthwhile now to compare the efficiency of the two methods
in general. In the first place, one does not reveal any secret
saying that CFD codes are computationally expensive. On the other
hand, having said that the scale of the gradients is one particle
diameter, one has served the trouble: Effectively, the cell size
cannot be  greater than a few particles. At values of the Mach
number of about 10 in some phases of the motion, inertia is the
most prominent mechanism of transport. Diffusion operates at very
small length scales, and is of less importance than inertia, but
plays a role in some other phases of the motion, when the typical
Mach number drops below 1 in the dense phase, and cannot be just neglected. In
addition, explicit time schemes like the one used in this paper
introduce a stability condition which turns more and more
restrictive as the diffusion length scales (the size of the
particles) become smaller. The result is that for submillimetric
particles, most commonly used in real experiments,  the same problem treated
here turns unbearably stiff, as the typical timescale of the pattern does not
decrease in the same correspondence.     

In any case one must resolve length scales of the typical size of
one particle, and the ratio one cell $\approx$ one particle is
extremely unfavorable to computational hydrodynamics in terms of
efficiency. Moreover, high resolution shock-capturing schemes like
WENO add an extra cost in the computation which often amounts the
largest portion of the total CPU time. There is some room for
optimization though, in the sense that not everywhere, not all the
time, such extraordinary effort to construct highly accurate
numerical fluxes should be necessary, and thus the simulation time could
this way be reduced.

The question about boundary conditions still deserves an
additional remark. Here the most simple choice has been made,
with specular reflecting boundaries without any portion of no-slip
in the tangential direction (the velocity component parallel to
the plate is not disturbed by the boundary). Presumably, in order
to mimic rough particles some partial slip would have to be
introduced with parameters to be determined from MD simulations,
although  friction may trigger rotational mechanisms which are not
contemplated in the set of hydrodynamic equations considered here.

Finally, one should ask why the agreement between molecular dynamics and
hydrodynamics is here so good, so robust, when all ingredients are present to
make it fail. At high densities collisions are not binary, fluxes are thus
nonlocal and the expressions for the kinetic coefficients should no longer
apply, not to mention the degree of inelasticity (the restitution is only
$e$ = 0.75) which invalidates the assumption of 
small gradients on which all kinetic theory expansions are based. It seems
that there is only one reason which can explain the agreement: The dynamics is
mostly Eulerian and details like constitutive relations  and transport
coefficients are not definitive. So much so that in practice, if one focuses
the attention on modeling accurately the shock wave and introduces a sensible
equation of state, the instability is ``easily'' reproduced, as in molecular
dynamics, with the essential contribution of inelasticity. 

\begin{acknowledgments}
JAC acknowledges the support from DGI-MEC (Spain) project
MTM2005-08024. The research of TP was supported by a Grant from
the G.I.F., the German-Israeli Foundation for Scientific Research
and Development. CS thanks support from project MEC:DPI2003-06725-C02-01 of the
Spanish Ministry of Education and Science. 
\end{acknowledgments}

\appendix

\section{Formulas for local variables}

Here, we summarize the formulas for the explicit eigenvalues and
eigenvectors of the Jacobian matrix of the Euler part of
\eqref{nscons} in 2D, for general equations of state depending on
the density and the enthalpy. As discussed in Section \ref{NS},
the system \eqref{nscons} can be rewritten as
\begin{equation}
  \frac{\partial \bu}{\partial t} + \frac{\partial}{\partial x_1}
\bff\left(\bu\right) + \frac{\partial}{\partial x_2}  \bg\left(\bu\right) = \bS\left(\bu\right)
\end{equation}
with obvious definitions for $\bff\left(\bu\right)$,
$\bg\left(\bu\right)$ and $\bS\left(\bu\right)$ and with the
vector $\bu = \left(u_1,u_2,u_3,u_4\right)$ whose components
are given by
\begin{equation}
  \begin{split}
    u_1 & =\rho \\ %\qquad , \qquad
    u_2 &=\rho U_1 \\ % \qquad , \qquad
    u_3 &=\rho  U_2 \\
    u_4 &=\rho \epsilon + \frac12 \rho \left|\bU\right|^2=\rho \epsilon + \frac12 \rho \left(U_1^2+U_2^2\right)\,.
  \end{split}
\end{equation}
The components of
$f$ are
\begin{equation}
  \begin{array}{lcl}
    \displaystyle f_{1}\left(\bu\right) =u_2 & , &
    \displaystyle f_{2}\left(\bu\right) =p+\frac{u_2^2}{u_1}\\
    \displaystyle f_{3}\left(\bu\right) = \frac{u_2 u_3}{u_1} & , &
    \displaystyle f_{4}\left(\bu\right) = \frac{u_2}{u_1} \left(p+u_4\right)
  \end{array}
\end{equation}
where the pressure is assumed to be a function of density and enthalpy,
\begin{equation}
  p=p\left(\rho,\epsilon\right)=p\left( u_1 , \frac{u_3}{u_1} - \frac12 \frac{u_2^2}{u_1^2}  \right)\,.
\end{equation}
The sound speed is given by
\begin{equation}
c_s^2 = \frac{\partial p}{\partial \rho} + \frac{p}{\rho^2}
\frac{\partial p}{\partial \epsilon} = p_\rho + \frac{p}{\rho^2}
p_\epsilon, \label{soundspeed}
\end{equation}
Given the auxiliary function is now $H= \left(p+u_4\right)/u_1$,
the eigenvalues of the Jacobian matrix
$\bff^\prime\left(\bu\right)$ are given by
\begin{equation}
    \Lambda_- = \frac{u_2}{u_1} - c_s\qquad , \qquad
    \Lambda_2 = \Lambda_3 = \frac{u_2}{u_1}\qquad \mbox{and} \qquad
    \Lambda_+ = \frac{u_2}{u_1} + c_s
\end{equation}
and their corresponding right and left eigenvectors are
\begin{equation}
  r_{\pm}=
  \begin{pmatrix}
    1 \\[1mm]  \frac{u_2}{u_1} \pm c_s \\[1mm] \frac{u_3}{u_1} \\[1mm] H \pm \frac{u_2}{u_1} c_s
  \end{pmatrix}
  \quad
  r_2=
  \begin{pmatrix}
    1 \\[1mm] \frac{u_2}{u_1} \\[1mm] \frac{u_3}{u_1} \\[1mm] H - \frac{1}{b_1}
  \end{pmatrix}
  \quad
  r_3=
  \begin{pmatrix}
    0 \\[1mm] 0 \\[1mm] 1 \\[1mm] \frac{u_3}{u_1}
  \end{pmatrix}
\end{equation}
and
\begin{equation}
l_{\pm}=
\begin{pmatrix}
  \frac{1}{2} \mp \frac{u_2}{u_1} \frac{1}{2c_s} + \frac{b_1}{2} \left( \frac{u_2^2 + u_3^2}{u_1^2} - H \right) \\[2mm]
  - \frac{b_1}{2} \frac{u_2}{u_1} \pm \frac{1}{2c_s} \\[2mm]
  - \frac{b_1}{2} \frac{u_3}{u_1} \\[2mm]
  \frac{b_1}{2}
\end{pmatrix}\,;~
l_2\!=\!
\begin{pmatrix}
  \left[ H-\frac{u_2^2}{u_1^2}-\frac{u_3^2}{u_1^2}  \right] b_1 \\[2mm]
  \frac{u_2}{u_1} b_1 \\[2mm]
  \frac{u_3}{u_1} b_1 \\[2mm]
  -b_1
\end{pmatrix}\;;~
l_3\!=\!
\begin{pmatrix}
  -\frac{u_3}{u_1} \\[2mm] 0 \\[2mm] 1 \\[2mm] 0
\end{pmatrix}
\end{equation}
respectively, where $b_1 = p_\epsilon\left/\rho c_s^2\right.$.
For the fluxes associated with the $x_2$-derivative, that is,
\begin{equation}
  \begin{array}{lcl}
    \displaystyle g_{1}\left(\bu\right) =u_3 & , &
    \displaystyle g_{2}\left(\bu\right) = \frac{u_2 u_3}{u_1}\\
    \displaystyle g_{3}\left(\bu\right) = p+\frac{u_3^2}{u_1} & , &
    \displaystyle g_{4}\left(\bu\right) = \frac{u_3}{u_1}
    \left(p+u_4\right);
  \end{array}
\end{equation}
we just need to swap indices 2 and 3 in all formulas above, and
the second and third components in all right and left
eigenvectors.

\bibliographystyle{jfm}
%\bibliography{cps_jfm}

\end{document}